\pdfoutput=1
\documentclass[usenatbib,ctexart]{mnras}

\usepackage[T1]{fontenc}
\usepackage{ae,aecompl}

\usepackage{graphicx}	
\usepackage{amsmath}	
\usepackage{amssymb}	
\usepackage{color}
\usepackage{aas_macros}
\usepackage{hyperref,breakurl}
\usepackage[usenames,dvipsnames]{xcolor}
\usepackage{dcolumn}
\usepackage{bm}
\usepackage{hyperref}
\usepackage{array}
\usepackage{dcolumn}
\usepackage{subfig}
\usepackage{amsmath,amssymb,latexsym,times}
\usepackage{float}
\usepackage{fontawesome5}
\usepackage{xurl}

\usepackage{eso-pic}

\AddToShipoutPictureBG*{%
  \AtPageUpperLeft{%
    \hspace{0.72\paperwidth}%
    \raisebox{-4.5\baselineskip}{%
      \makebox[0pt][l]{\textnormal{DES-2019-0504}}
}}}%

\AddToShipoutPictureBG*{%
  \AtPageUpperLeft{%
    \hspace{0.72\paperwidth}%
    \raisebox{-5.5\baselineskip}{%
      \makebox[0pt][l]{\textnormal{FERMILAB-PUB-23-086-PPD}}
}}}%

\usepackage{etoolbox}
\pretocmd{\abstractname}{\newpage}{}{}

\usepackage{lineno}
\usepackage{newtxtext,newtxmath}

\newcommand{\cwr}[1]{#1}

\newcommand{\hinv}{h^{-1}}
\newcommand{\degree}{^{\circ}}

\newcommand{\updated}[1]{#1}

\title[Intrinsic Alignment in DES Y1 redMaPPer Clusters]{The Intrinsic Alignment of Red Galaxies in DES Y1 redMaPPer Galaxy Clusters}

\author[Zhou \& Tong et al.]{
\parbox{\textwidth}{
\Large
C.~Zhou,$^{1,2}$\thanks{Email: zhou.conghao@ucsc.edu}
A.~Tong,$^{1}$\thanks{Email: alexander.tong@duke.edu}
M.~A.~Troxel,$^{1}$
J.~Blazek,$^{3}$
C.~Lin,$^{1}$
D.~Bacon,$^{4}$
L.~Bleem,$^{5}$
C.~Chang,$^{6,7}$
M.~Costanzi,$^{8,9,10}$
J.~DeRose,$^{11}$
J.~P.~Dietrich,$^{12}$
A.~Drlica-Wagner,$^{6,13,7}$
D.~Gruen,$^{12}$
R.~A.~Gruendl,$^{14,15}$
B.~Hoyle,$^{12}$
M.~Jarvis,$^{16}$
N.~MacCrann,$^{17}$
B.~Mawdsley,$^{4}$
T.~McClintock,$^{18}$
P.~Melchior,$^{19}$
J.~Prat,$^{6,7}$
A.~Pujol,$^{20,21}$
E.~Rozo,$^{18}$
E.~S.~Rykoff,$^{22,23}$
S.~Samuroff,$^{3}$
E.~Sheldon,$^{24}$
T.~Shin,$^{25}$
A.~Carnero~Rosell,$^{26,27,28}$
B.~Yanny,$^{13}$
C.~S{\'a}nchez,$^{16}$
D.~L.~Tucker,$^{13}$
I.~Sevilla-Noarbe,$^{29}$
J.~Zuntz,$^{30}$
T.~N.~Varga,$^{31,32,33}$
Y.~Zhang,$^{34}$
O.~Alves,$^{35}$
A.~Amon,$^{36,37}$
E.~Bertin,$^{38,39}$
D.~Brooks,$^{40}$
D.~L.~Burke,$^{22,23}$
M.~Carrasco~Kind,$^{14,15}$
L.~N.~da Costa,$^{27}$
T.~M.~Davis,$^{41}$
J.~De~Vicente,$^{29}$
S.~Desai,$^{42}$
H.~T.~Diehl,$^{13}$
P.~Doel,$^{40}$
S.~Everett,$^{43}$
I.~Ferrero,$^{44}$
B.~Flaugher,$^{13}$
J.~Frieman,$^{13,7}$
D.~W.~Gerdes,$^{34,35}$
G.~Gutierrez,$^{13}$
S.~R.~Hinton,$^{41}$
D.~L.~Hollowood,$^{45}$
K.~Honscheid,$^{46,47}$
D.~J.~James,$^{48}$
T.~Jeltema,$^{45}$
K.~Kuehn,$^{49,50}$
O.~Lahav,$^{40}$
M.~Lima,$^{51,27}$
J.~L.~Marshall,$^{52}$
J. Mena-Fern{\'a}ndez,$^{29}$
F.~Menanteau,$^{14,15}$
R.~Miquel,$^{53,54}$
A.~Palmese,$^{55}$
F.~Paz-Chinch\'{o}n,$^{14,36}$
A.~Pieres,$^{27,56}$
A.~A.~Plazas~Malag\'on,$^{19}$
A.~Porredon,$^{46,47,30}$
M.~Raveri,$^{57}$
A.~K.~Romer,$^{58}$
E.~Sanchez,$^{29}$
M.~Smith,$^{59}$
M.~Soares-Santos,$^{35}$
E.~Suchyta,$^{60}$
M.~E.~C.~Swanson,$^{61}$
G.~Tarle,$^{35}$
C.~To,$^{46}$
N.~Weaverdyck,$^{35,11}$
J.~Weller,$^{32,33}$
and P.~Wiseman$^{59}$
\begin{center} (DES Collaboration) \end{center}
}
\vspace{-0.9cm}}

\date{Accepted XXX. Received YYY; in original form ZZZ}

\pubyear{2023}

\begin{document}
\label{firstpage}
\pagerange{\pageref{firstpage}--\pageref{lastpage}}
\maketitle

\begin{abstract}
    Clusters of galaxies trace the most nonlinear peaks in the cosmic density field. The weak gravitational lensing of background galaxies by clusters can allow us to infer their masses. However, galaxies associated with the local environment of the cluster can also be intrinsically aligned due to the local tidal gradient, contaminating any cosmology derived from the lensing signal. We measure this intrinsic alignment in Dark Energy Survey (DES) Year 1 \textsc{redMaPPer} clusters. We find evidence of a non-zero mean radial alignment of galaxies within clusters between redshift $0.1-0.7$. We find a significant systematic in the measured ellipticities of cluster satellite galaxies that we attribute to the central galaxy flux and other intracluster light. We attempt to correct this signal, and fit a simple model for intrinsic alignment amplitude ($A_{\textrm{IA}}$) to the measurement, finding $A_{\textrm{IA}}=0.15\pm 0.04$, when excluding data near the edge of the cluster. We find a significantly stronger alignment of the central galaxy with the cluster dark matter halo at low redshift and with higher richness and central galaxy absolute magnitude (proxies for cluster mass). This is an important demonstration of the ability of large photometric data sets like DES to provide direct constraints on the intrinsic alignment of galaxies within clusters. These measurements can inform improvements to small-scale modeling and simulation of the intrinsic alignment of galaxies to help improve the separation of the intrinsic alignment signal in weak lensing studies.
\end{abstract}

\begin{keywords}
    cosmology: observations -- gravitational lensing: weak -- galaxies: clusters: general
\end{keywords}






\section{Introduction}
In 1919, predictions from the theory of general relativity were confirmed by observing the deflection of the light by the sun \citep{10.2307/91137}, which is aptly named gravitational lensing. A century after this experiment, gravitational lensing has become one of the most powerful probes in modern cosmology surveys. Weak lensing probes including galaxy-galaxy lensing, cluster lensing, and cosmic shear can effectively constrain cosmological parameters and thus reveal the growth history of structure in the universe. The recent growth in data volume from Stage III surveys such as the Dark Energy Survey (DES),\footnote{\url{https://www.darkenergysurvey.org}} the Kilo-Degree Survey,\footnote{\url{https://kids.strw.leidenuniv.nl}} and the Hyper Suprime-Cam Survey\footnote{\url{https://hsc.mtk.nao.ac.jp/ssp/}} has significantly lowered the statistical uncertainty in the lensing signal. This has in turn made control of small systematic errors critical for extracting weak lensing signals from existing and future surveys.

One major source of systematic uncertainty in weak lensing studies is from the correlated intrinsic alignment of galaxies that contaminate the shear correlations \citep{troxelreview}. The intrinsic alignment of galaxies is caused by a variety of physical processes during structure formation \cite{2000MNRAS.319..649H,2000ApJ...545..561C,hirata04,bridle07,2019PhRvD.100j3506B}, leading to a tendency for galaxies to physically align along the gradient of the tidal field. The intrinsic alignment of galaxies acts as a nuisance signal to the lensing measurement, which tends to distort the observed shape of a galaxy tangentially to the gradient of the tidal field, and it can strongly bias the weak lensing results we infer (e.g., \cite{2019PhRvD.100j3506B,2020PASJ...72...16H,2021A&A...645A.104A,2021arXiv210513548K,2022PhRvD.105b3520A}) if it is improperly corrected or modeled. Isolating the intrinsic alignment signal can not only improve the results we get from lensing surveys, but also provides insights into the evolution of galaxies over time, which would also modify the intrinsic alignment signal.

The alignment of galaxies in large-scale tidal fields has been well studied and especially for large and red galaxies, there is a consensus in both measurements and simulations that a non-zero alignment exists (e.g., \cite{2006MNRAS.367..611M,
    2007MNRAS.381.1197H,
    2011A&A...527A..26J,
    2013MNRAS.436..819J,
    2015MNRAS.454.2736C,
    2015MNRAS.450.2195S,
    2016MNRAS.462.2668T,
    2019MNRAS.489.5453S,
    2021A&A...654A..76F,
    2021MNRAS.508..637S,
    2020arXiv201007951Z}). Ignoring destructive interference via interaction or merging of galaxies and clusters, one naively expects that the intrinsic galaxy alignment would be stronger around the strongest over-densities in the universe like galaxy clusters. There is more disagreement about the amplitude of the alignment of galaxies within such large structures, i.e., intracluster alignments (e.g., \cite{2005ApJ...627L..21P,
    2006ApJ...644L..25A,
    2007ApJ...662L..71F,
    2009arXiv0903.2264S,
    2011ApJ...740...39H,
    2013MNRAS.433.2727S,
    2015A&A...575A..48S}) with different shape measurement methods leading to different conclusions. A measurement of the alignment of \textsc{redMaPPer} clusters in the Sloan Digital Sky Survey (SDSS) data with the large-scale matter field was also performed by \cite{2017MNRAS.468.4502V}. \cite{Huang_2017} also found that the inferred alignment depended also on the population of galaxies, which may inform discrepancies among earlier studies.

The substantially increased physical volume (and thus the number of clusters) probed in data sets like the Dark Energy Survey Year 1 data enable an extremely powerful test of this question of intracluster alignment. In this work, we study a variety of alignment mechanisms for red-sequence galaxies within DES Year 1 \textsc{redMaPPer} clusters. This follows an earlier work studying \textsc{redMaPPer} clusters in the Sloan Digital Sky Survey (SDSS) data \citep{Huang_2016,Huang_2017}. We examine a similar set of alignment statistics as this earlier work, comparing the \textsc{metacalibration} and \textsc{im3shape} weak lensing shape measurement algorithms used in DES Year 1 for cosmology. In particular, we are able to measure a significant non-zero signal in the metric most of interest to cosmology, the mean tangential (radial) shear. These measurements demonstrate that current and future large photometric surveys are able to provide significant constraints on these local alignment processes.

The paper is organized as follows. In Sec. \ref{data} we discuss the DES data used in this work, including the cluster and shape catalogs. We describe the methodology used in Sec. \ref{methods}, and the measurement results in Sec. \ref{results}. In Sec. \ref{model} we present a discussion of the interpretation of the signal in terms of an intrinsic alignment model and the mass profiles of the clusters. We conclude in Sec. \ref{conclusions}.

\section{Dark Energy Survey Year 1 Data}\label{data}
The Dark Energy Survey is a six-year survey covering 5000 square degrees of the southern sky using the Dark Energy Camera \citep{decam} mounted on the Blanco 4m telescope in Cerro Tololo, Chile. Observations use five broadband filters $g, r, i, z , Y$. The first year of DES observations (Y1) lasted from August 2013 to February 2014 and covers $\sim$40\% of the total DES footprint \citep{y1gold}. We use data based on several value-added catalogs built from the Y1 data: 1) the Y1A1 GOLD catalog, a high-quality photometric data set; 2) the red-sequence Matched-filter Probabilistic Percolation (\textsc{redMaPPer}) cluster and member catalogs; 3) the \textsc{metacalibration} and \textsc{im3shape} shape catalogs. We describe each of these in more detail in the following subsections.

\subsection{GOLD Catalog}
\label{sec:y1a1gold}
The Y1A1 GOLD data set \citep{y1gold} is a high-quality photometric catalog that contains multi-epoch, multi-object photometric model parameters, and other ancillary information. The objects in this catalog are selected from the initial Y1A1 coadd detection catalog, which is processed by the DESDM image processing pipeline \citep{2011arXiv1109.6741S,2008SPIE.7016E..0LM,2012SPIE.8451E..0DM}. The Y1A1 GOLD catalog restricts the footprint of the objects to regions with at least one image of sufficient science quality in each filter. Several bad region masks including unphysical colors, the Large Magellanic Cloud, globular clusters, and bright stars are applied to the catalog. The final Y1A1 GOLD footprint covers $\sim$1800 deg$^2$ with an average of three to four single-epoch images per band. The photometric accuracy is \(\lesssim 2 \%\) over the survey area. A comparison with the deeper catalog of the Canada-France-Hawaii Telescope Lensing Survey shows that the Y1A1 GOLD catalog is $>$ 99\% complete in $g,r,i,z$ bands for magnitudes brighter than 21.5. There are approx. 137 million objects in the final Y1A1 GOLD catalog.

\subsection{\textsc{redMaPPer} cluster catalog}
\label{sec:redMaPPer} 

The red-sequence Matched-filter Probabilistic Percolation (\textsc{redMaPPer}) photometric cluster finding algorithm is optimized for deep wide-field photometric cosmology surveys \citep{2014ApJ...785..104R} and produces a cluster catalog identifying overdensities of red-sequence galaxies with a probabilistic assignment of these red-sequence galaxies as central/satellite members. This alogorithm has been validated using X-ray and Sunyaev-Zel'dovich (SZ) observations \citep{2015MNRAS.453...38R,2015MNRAS.454.2305S,2016MNRAS.461.1431R,2014A&A...571A..87S,2020ApJS..247...25B,2021MNRAS.504.1253G}, and updates to the method are described in \cite{2016MNRAS.461.1431R,2016ApJS..224....1R,2019MNRAS.482.1352M}. We briefly describe the algorithm and resulting cluster catalog below.

To identify clusters, the \textsc{redMaPPer} algorithm counts the excess number of red-sequence galaxies, called the richness ($\lambda$), within a radius $R_\lambda = 1.0  \hinv  \textrm{Mpc} (\lambda/100)^{0.2}$ that are brighter than some luminosity threshold $L_{\mathrm{min}}(z)$. A locally volume-limited version of the catalog is also produced, which imposes a maximum redshift on clusters such that galaxies above $L_{\mathrm{min}}(z)$ can be detected at 10$\sigma$. An associated redshift-dependent random catalog for both cluster catalogs is produced using a survey mask constructed to require that a cluster at redshift $z$ at each point in the mask be masked by at most 20\% by the associated galaxy footprint mask.

The algorithm centers each cluster on the most likely central galaxy, based on an iteratively-trained filter relying on galaxy brightness, cluster richness, and local density to determine the central candidate probability. Each red-sequence cluster member is also assigned an associated membership probability, which we weigh all measurements by. Additional information about the quality of photometric redshifts of the clusters and cluster members can be found in \cite{2019MNRAS.482.1352M,wthetapaper}, but over most of the redshift range used in this paper cluster redshifts are unbiased at the level of $|\Delta z| \leq 0.003$ with a median photometric redshift scatter of $\sigma_z/(1+z)\approx 0.006$. For red-sequence cluster members, this is $\sigma_z/(1+z)\approx 0.035$.

\begin{figure}
    \begin{center}
        \includegraphics[width=\columnwidth]{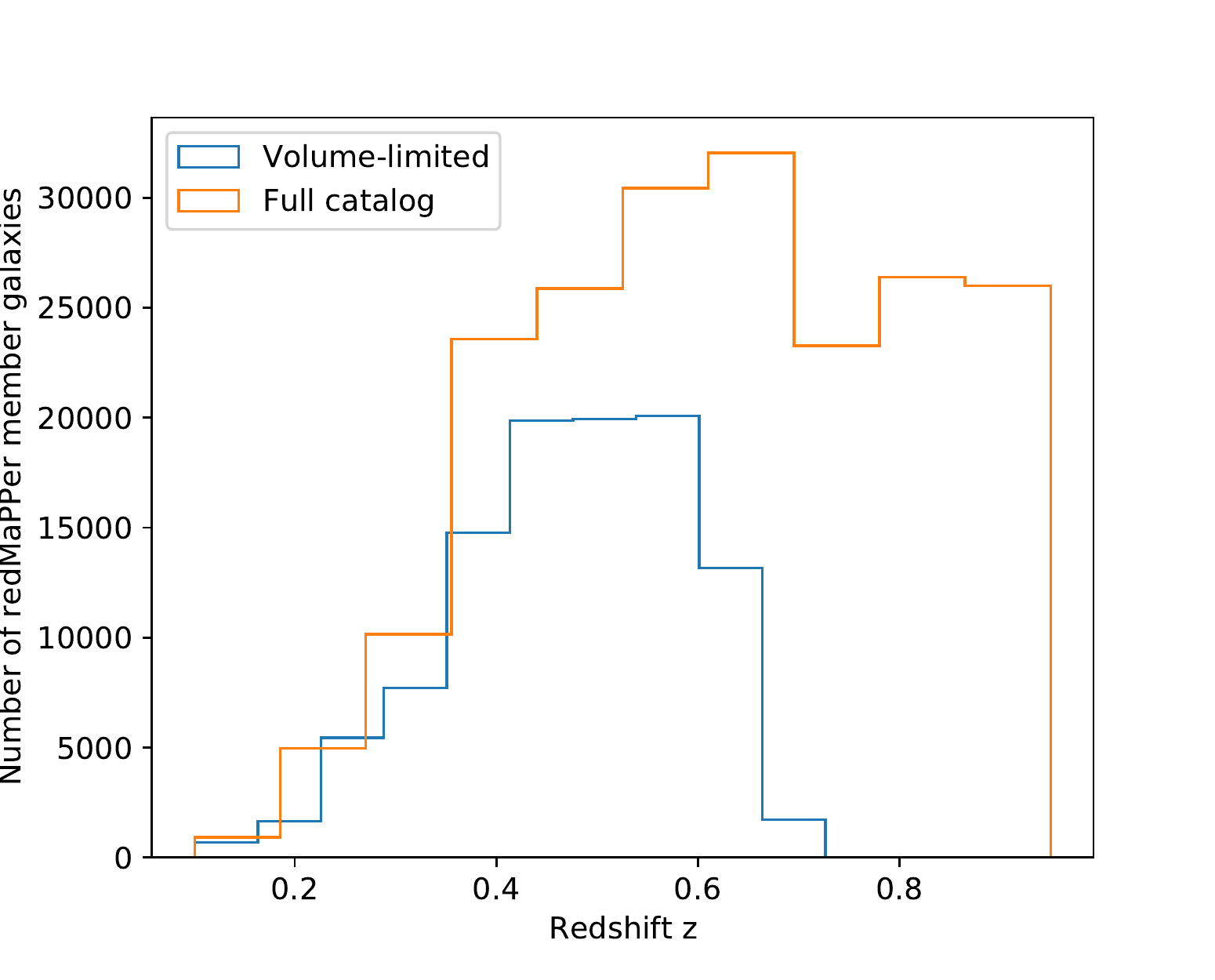}
    \end{center}
    \caption[]{The redshift distribution of \textsc{redMaPPer} cluster members used in this work.
        \label{fig:zdist}}
\end{figure}

In this work, we use a total of 16966 clusters from the DES Y1 \textsc{redMaPPer} catalog (7066 in the volume-limited catalog). Within these clusters, there are an effective number of 452280 (248670) cluster members (either central or satellite galaxies). We have performed measurements both using all clusters and only the volume-limited sample. The full catalog allows us to probe a larger redshift range with higher statistical precision, while the volume-limited sample matches what has been used for cosmological inference in \cite{2020arXiv200211124D}. We will show results primarily from the volume-limited sample unless otherwise noted for cases where results are not qualitatively similar, and using the same $\lambda>20$ selection on richness in either case as \cite{2020arXiv200211124D}, since inference of the halo shape based on the distribution of satellite galaxies is increasingly difficult as the number of satellite galaxies decreases. We use as central galaxies only the most probable central galaxy in each cluster. The redshift distributions of the final samples of clusters are shown in Fig. \ref{fig:zdist}.     \updated{It is important to note that the full catalog, relative to a volume-limited selection, will have some systematic selection bias in the population of clusters probed, particularly at around $z=0.7$ and above. Characterizing this selection is beyond the scope of this paper.}

\subsection{Shape Catalogs}

We use a fiducial shape catalog that is calibrated with the \textsc{metacalibration} method, which uses available imaging data directly without the need for significant prior information as a function of galaxy properties  \citep{HuffMandelbaum2017,SheldonHuff2017}. The \textsc{metacalibration} implementation used in DES Y1 was described in detail in \cite{shearcat}. Limitations in the DES Y1 implementation of \textsc{metacalibration} lead to a residual mean multiplicative shear bias estimate of $m = 0.012 \pm 0.013$, which is due primarily to the effects of neighboring light on the shear recovery. This mean correction is applied to the measurements in this work. For \textsc{im3shape}, we divide the mean shear signal by the mean of $1+m$, where $m$ is the calibration factor inferred from simulations, and for \textsc{metacalibration}, we divide the mean shear signal by the mean value of $\frac{1}{2}(1+m)(\boldsymbol{R}_{11} + \boldsymbol{R}_{22}$), where $m$ is the shear bias estimate above and $R$ the response inferred from the \textsc{metacalibration} process.

    \textsc{metacalibration} also allows us to account for sample selection bias effects, as described in \cite{shearcat,shearcorr}, which we also include. However, we match the shape catalog to the \textsc{redMaPPer} central/satellite member catalog, which introduces an additional selection that we cannot incorporate in the selection bias correction. In future work, it would be valuable to explore the impact of this selection by running the \textsc{redMaPPer} selection algorithm on the photometry produced in the \textsc{metacalibration} process similar to how we incorporate redshift selection biases in, e.g., \cite{shearcorr}. This has been measured, for example, for a generic red galaxy selection used for intrinsic alignment studies in \cite{2019MNRAS.489.5453S}. At the current precision of the measurements in this paper, however, we expect this additional correction to be safely negligible.
    The \textsc{metacalibration} catalog yields a total of 35 million objects, 262867 of which are matched to the \textsc{redMaPPer} central/satellite members and used in the selection for the current analysis. We are able to match a \textsc{metacalibration} shape measurement to 66\% of \textsc{redMaPPer} members.

    We also compare measurements using the \textsc{im3shape} shape catalog \cite{shearcat,2013MNRAS.434.1604Z}, which utilizes a simulation-based calibration and only has secure shape measurements for 39\% of \textsc{redMaPPer} members. This low fraction of cluster members with secure shapes for \textsc{im3shape} gives too low a signal-to-noise for the two-point correlation function measurements presented later in Sec. \ref{radial} to be useful, but it is compared to \textsc{metacalibration} in other measurements. The \textsc{im3shape} catalog provides a model fit for either a bulge- or disk-like profile. We find about 80\% of central galaxies better fit by a de Vaucouleurs (bulge) profile vs exponential (disk) profile, while for satellites, about 60\% are better fit by an exponential profile.


    \section{Methods to infer the intrinsic alignment of galaxies in clusters}\label{methods}

    The intrinsic alignment of galaxies in the (quasi-)linear regime is typically expressed via perturbation theory as a function of the underlying tidal field. Most cosmological studies have used a linear alignment model \citep{hirata04,bridle07} that uses the first-order expansion of the intrinsic shear $\gamma^I$ (shown here up to second-order) in the linear density field:
    \begin{equation}
        \gamma^I(\bm{x}) = C_1 s_{ij}+C_2\left(s_{ik}s_{kj}-\frac{1}{3}\delta_{ij}s^2\right)+C_{1\delta}(\delta s_{ij})+C_t t_{ij}+\cdots,
    \end{equation}
    where each field is evaluated at $\bm{x}$ and summation occurs over repeated indices. The $C_i$ parameters are then the analog to galaxy bias parameters in perturbation theory, and $\delta_{ij}$ is the Kronecker delta, $\delta$ is the density field, $s_{ij}(\bm{k})\equiv\hat{S}_{ij}[\delta(k)]$ is the normalized Fourier-space tidal tensor, $s^2(\bm{k})$ is the tidal tensor squared, and the tensor $t_{ij}=\hat{S}_{ij}[\theta-\delta]$ involves the velocity shear. From this, one can build up all standard components of commonly used intrinsic alignment models up to second order in the density field, as described in detail in \cite{2019PhRvD.100j3506B}.

    When modeling the intrinsic alignment of galaxies in strongly nonlinear environments like galaxy clusters, where perturbative models will break down, it has been proposed to use a `1-halo' model in analogy to the halo model for the matter power spectrum to describe alignments internal to a single cluster halo. This has been discussed by \cite{SB10, 2021MNRAS.501.2983F}, which outlines approaches for building such a model, including tests on simulations. Previous attempts to directly measure such a signal, e.g. within galaxy clusters, have had mixed results both in simulations and data. These fall into two categories: 1) the alignment of the cluster shape with the tidal field and 2) the alignment of satellite galaxies, using the cluster centers as a proxy for the peaks of the local tidal field.

    Better measurements of the 1-halo intrinsic alignment signal are necessary to inform and constrain such a beyond-perturbative model, however, which is the goal of this paper. While most measurement attempts have focused on objects with spectroscopic redshifts, which suffer from limited data volumes, we present several complementary measurements of these alignments using a fully photometric galaxy cluster and satellite catalog that selects red-sequence galaxies and spans over 1000 deg$^2$ to redshift 0.7.

    \subsection{Orientation of the satellite galaxy distribution}\label{pamethod}

    \begin{figure}
        \begin{center}
            \includegraphics[width=0.9\columnwidth]{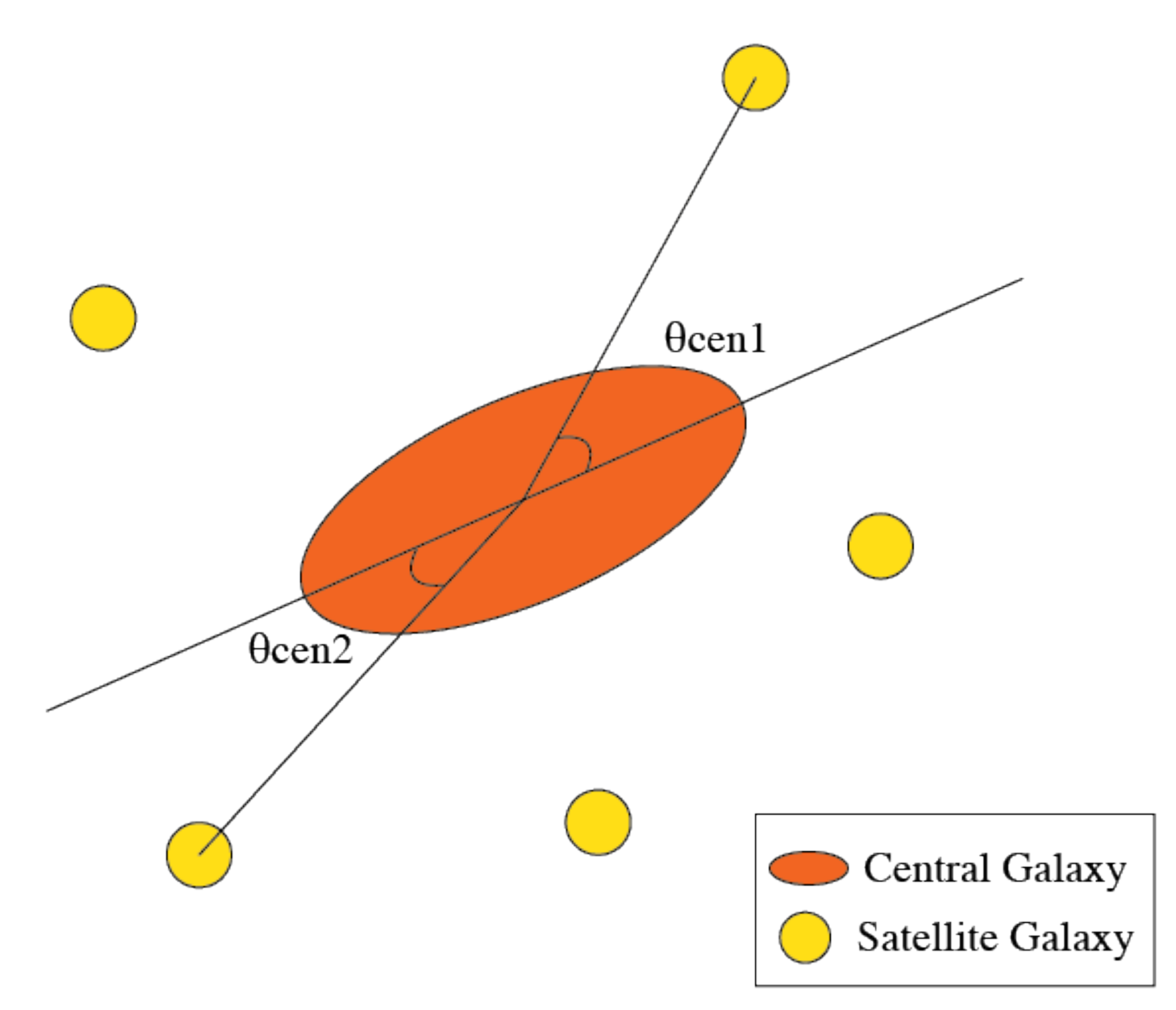}
            \includegraphics[width=\columnwidth]{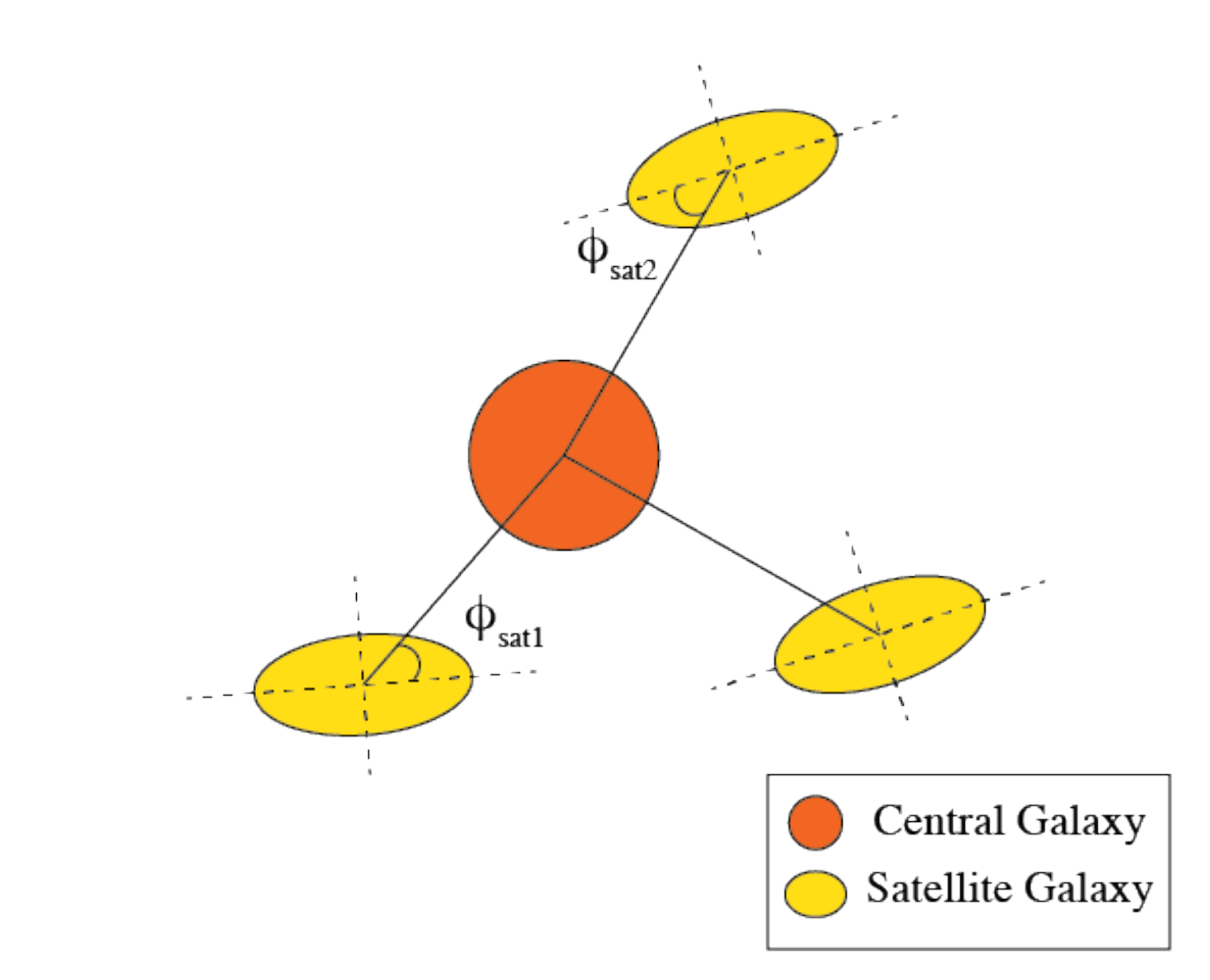}
        \end{center}
        \caption[]{\emph{Top}: Measured quantities relevant to the orientation of the central galaxy within the dark matter halo of the cluster. $\Delta \eta$ is the position angle difference between the central galaxy and the cluster halo. $\theta_{\textrm{cen}}$ is the alignment angle of the line connecting the central galaxy and each satellite galaxy relative to the central galaxy position angle. \emph{Bottom}: Measured quantities relevant to the orientation of the satellite galaxies within the dark matter halo of the cluster. $\phi_{\textrm{sat}}$ is the alignment angle of the line connecting the central galaxy and each satellite galaxy relative to the satellite galaxy position angle.
            \label{fig:centralangle}}
    \end{figure}

    We quantify the strength of the central galaxy alignment relative to the orientation of the cluster satellite distribution as a proxy for the dark matter halo orientation in two ways, which were also used in SDSS for \textsc{redMaPPer} clusters by \cite{Huang_2016}. First, we use the position angle difference $\Delta \eta$ between the central galaxy and its host cluster, and second, the central galaxy alignment angle  $\theta _{cen}$ for each central-satellite pair. They are both defined to lie in the range $[0 \degree, 90 \degree]$, with values closer to $0 \degree$ indicating stronger central galaxy alignment.

    Measuring $\Delta \eta$ requires an approximation of the overall cluster shape from the distribution of satellite galaxies. We use 2 different methods to determine the ellipticity and orientation of the cluster in order to measure $\Delta \eta$. 
    \updated{Both measurements are most sensitive to the ellipticity at a range of radii close to half the cluster scale identified by $R_{\lambda}$.}

    \subsubsection{Method 1: Second moments}

    We follow the method used by \citet{Huang_2016} to calculate the cluster ellipticity and position angle of the satellite galaxies with respect to the central galaxy. We use all satellite galaxies with $p_{\mathrm{mem}} \geq 0.2$\footnote{The choice of minimum $p_{\mathrm{mem}}$ is arbitrary, and has very little impact on our results.} in order to reasonably trace the shape of the cluster. We first calculate the reduced second moments of the positions of all remaining satellite galaxies in the cluster:
    \begin{align}
        M_{xx} & \equiv \frac{\sum_i p_{i,\mathrm{mem}} \frac{x_i^2}{r_i^2}}{\sum_i p_{i,\mathrm{mem}}}   \\
        M_{xy} & \equiv \frac{\sum_i p_{i,\mathrm{mem}} \frac{x_i y_i}{r_i^2}}{\sum_i p_{i,\mathrm{mem}}} \\
        M_{yy} & \equiv \frac{\sum_i p_{i,\mathrm{mem}} \frac{y_i^2}{r_i^2}}{\sum_i p_{i,\mathrm{mem}}}
    \end{align}
    where $x_i$ and $y_i$ are the distances of satellite galaxy $i$ from the central galaxy in RA and Dec, respectively, and $r_i$ is the Cartesian distance from satellite galaxy $i$ to the central galaxy. We then use the Stokes parameters to define the cluster shape as follows:
    \begin{equation}
        (Q, U) = \frac{1-b^2/a^2}{1+b^2/a^2} (\cos 2\beta, \sin 2\beta) = (M_{xx} - M_{yy}, 2M_{xy})
    \end{equation}
    where $b/a$ is the cluster minor-to-major axis ratio and $\beta$ is the cluster position angle (PA).

    \subsubsection{Method 2: Quadrant grid}\label{qgsec}

    Our second method for measuring cluster shapes is based on the assumption that satellite projections are distributed isotropically along a profile of 2D ellipses around the central galaxy. We place a set of orthogonal axes on the central galaxy in the plane of the sky, rotated at different angles $\theta$ relative to the central galaxy position angle, and sum the $p_{mem}$ for all satellites in each quadrant ($q$).

\begin{figure}
        \begin{center}
            \includegraphics[angle=270,width=\columnwidth]{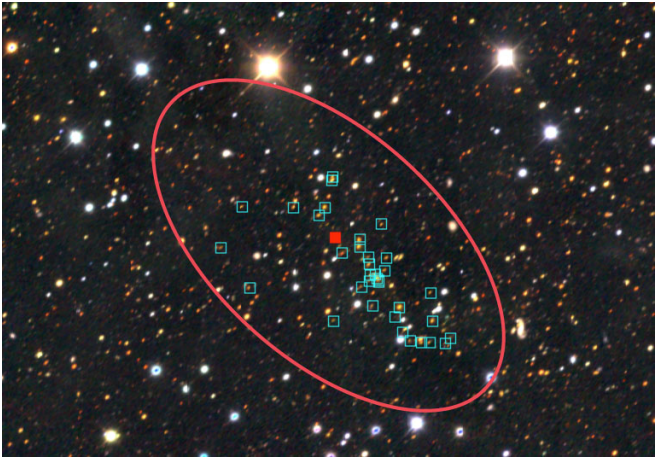}
        \end{center}
        \caption[]{An example \textsc{redMaPPer} identified cluster at $z=0.41$. Overlaid in red is the shape of the cluster fit by Method 2. This cluster was found to have $e=0.73$, with a position angle 48\textdegree~east-of-north and \textsc{redMaPPer} radius 0.746 Mpc. Member galaxies are identified in cyan squares to differentiate from other projected galaxies along the line-of-sight. The \cwr{brightest central galaxy} is the solid red square in the center. The model is constrained to be centered on the \textsc{redMaPPer}-identified central galaxy.
            \label{fig:cluster}}
    \end{figure}

    We define the count difference in cross-pair quadrants as $m = q_1 + q_3 - q_2 - q_4$, which we can model as a function of $\theta$. The assumption of a 2D ellipse leads to the following expression for $m(\theta)$:

    \begin{equation}
        m(\theta) = \frac{N}{2\pi}\left[ \arctan\left(\frac{\tan(\beta - \theta)}{r}\right) + 2 \arctan\left(\frac{\cot(\beta - \theta)}{r}\right) \right]
    \end{equation}
    where $N$ is the effective number of satellites in the cluster, $\beta$ is the cluster position angle, and $r$ is the minor-to-major axis ratio $b/a$. We fit this model to the count difference data as a function of $\theta$ assuming Poisson uncertainty and find the best-fit parameters $\beta$ and $r$, which together completely describe the shape of the cluster. An example cluster with the best-fit shape model over-plotted is shown in Fig. \ref{fig:cluster}.

    \begin{figure}
        \begin{center}
            \includegraphics[width=\columnwidth]{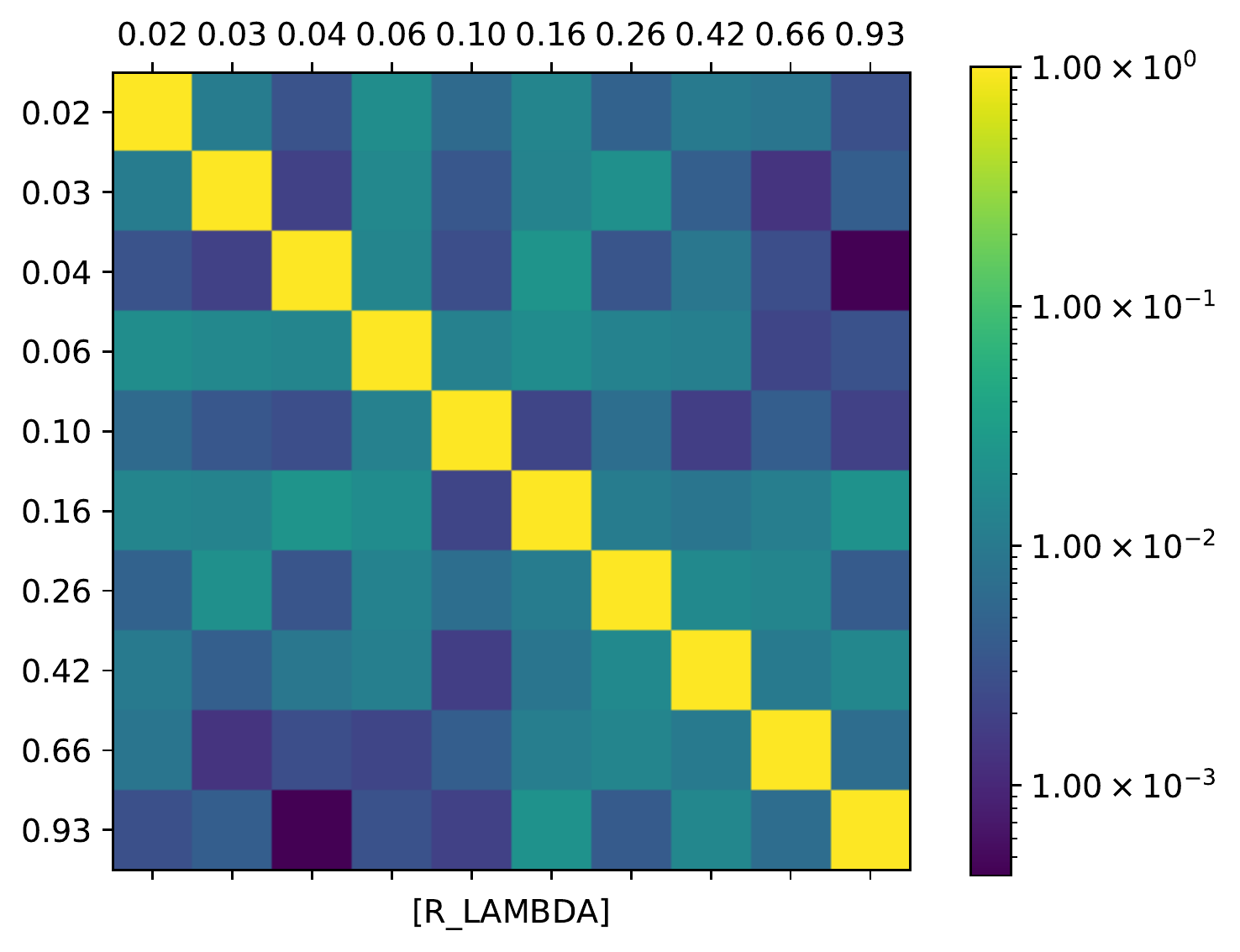}
        \end{center}
        \caption[]{The jackknife correlation matrix for the full-sample $\gamma_{T}(R)$ measurement, discussed in Secs. \ref{methodradial} \& \ref{radial}. As expected for shot or shape noise, the covariance is strongly diagonal.
            \label{fig:cov}}
    \end{figure}

    \subsection{Radial alignment of satellite galaxies with the cluster center}\label{methodradial}

    The tendency of satellite galaxies to align radially with their major axis pointed toward the central galaxy is another measure of the influence of the cluster's tidal field on the orientation of galaxies within its dark matter halo. While the mechanism for this alignment, e.g., whether it is achieved over time or during the galaxies' formation, is not clear, we can place empirical constraints on this alignment at the time we observe the cluster. We can then study the evolution of the mean alignment over time at different redshifts.

    One way to parameterize this alignment is similar to the observables described in the preceding section, which we will label $\phi_{\mathrm{sat}}$ following \cite{Huang_2017}. This is the angle between the position angle of the satellite galaxy and the line connecting it to the central galaxy. This is shown in Fig. \ref{fig:centralangle}.

    Another standard method is calculating the mean radial shape $\gamma_{T}(R)$
    \begin{equation}
        \gamma_{T}(R) = \frac{\sum_{i} p_{i,\mathrm{mem}} e_{i,+}}{\sum_{i} p_{i,\mathrm{mem}}}
    \end{equation}
    via the two-point correlation function of the central galaxy positions with the ellipticity of the satellite galaxies. $R$ is the projected distance separation of the satellite from the central galaxy of the cluster, $i$ is some satellite galaxy in some cluster, and $e_{+}$ is the component of the ellipticity projected along a basis coinciding with the line connecting the satellite galaxy to the central galaxy of the cluster. $\gamma_{T}$ is most relevant for contamination to the cluster lensing signal.
    In practice, we use TreeCorr\footnote{\url{https://github.com/rmjarvis/TreeCorr}} \citep{2004MNRAS.352..338J} to perform correlation function measurements in 10 logarithmic bins of the distance between the central galaxy and the satellite galaxies. The lower bound is arbitrary, while the upper bound is the maximum radial distance to a satellite galaxy.

    \begin{figure}
        \begin{center}
            \includegraphics[width=\columnwidth]{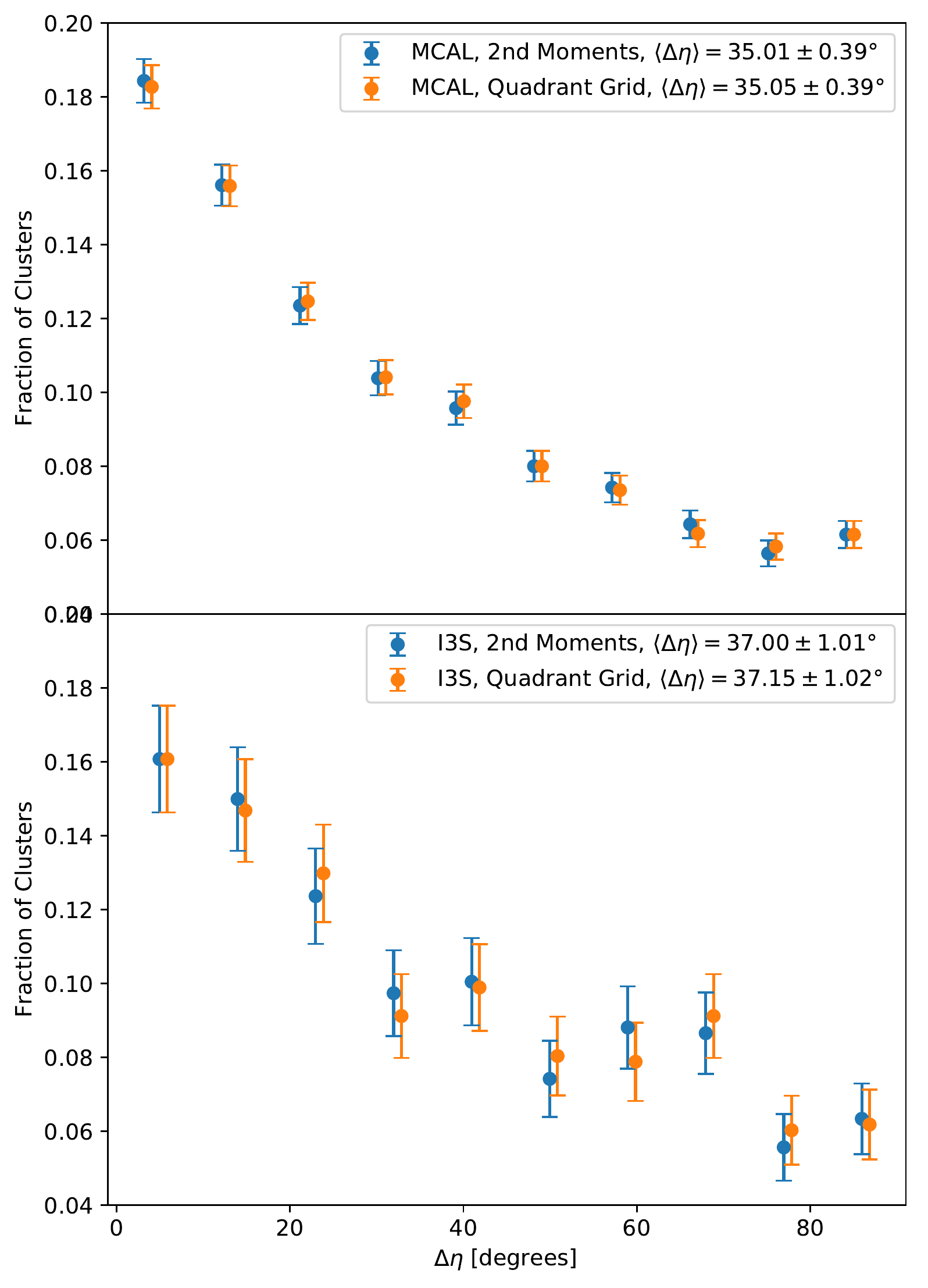}
        \end{center}
        \caption[]{The position angle differences ($\Delta \eta$) between the \cwr{brightest central galaxy} major axis and that of the satellite galaxy distribution for the DES Y1 galaxy clusters, as measured by the two methods described in Sec. \ref{pamethod}. \emph{Top}: The distribution with cluster \cwr{position angle} inferred from the \textsc{metacalibration} (MCAL) shape catalog. \emph{Bottom}: The distribution with cluster \cwr{position angle} inferred from the im3shape (I3S) catalog. The results are generally consistent with each other.
            \label{fig:padiff1}}
    \end{figure}

    \subsection{Estimating the covariance of measurements}

    Lacking a robust a priori theoretical model for what the measured signals should be, we cannot construct a theoretical covariance framework. Instead, we rely on a jackknife covariance estimate, iteratively removing each cluster from the sample. The covariance is then given by
    \begin{equation}
        C_{\xi}(x) = \frac{N-1}{N}\sum_{i=1}^N (\xi_i-\bar{\xi})^2,
    \end{equation}
    where $N$ is the number of clusters, $i$ is the cluster number, and $\bar{\xi}=\sum_i \xi_i/N$, for some estimator $\xi$. The covariances are expected to be dominated by shot or shape noise, given the small sample sizes, so we expect the jackknife approach to be sufficiently accurate. In particular, the measurement of $\gamma_{T}$ in Sec. \ref{radial}, which is the most substantial result in this work, is non-zero only for very small separations, where shape noise dominates the correlation function. The covariance matrix for $\gamma_{T}$ is shown in Fig. \ref{fig:cov}.

    \section{Measured alignment in DES clusters}\label{results}

    We present the results of the measurements described in the previous section. Unless otherwise noted, we will limit results to the volume-limited \textsc{redMaPPer} cluster catalog for brevity, since in most cases the results are qualitatively similar and thus conclusions drawn from the data will not differ.

    \subsection{Alignment of central galaxy with satellite galaxy distribution}\label{align1}

    \begin{figure*}
        \begin{center}
            \includegraphics[width=0.45\textwidth]{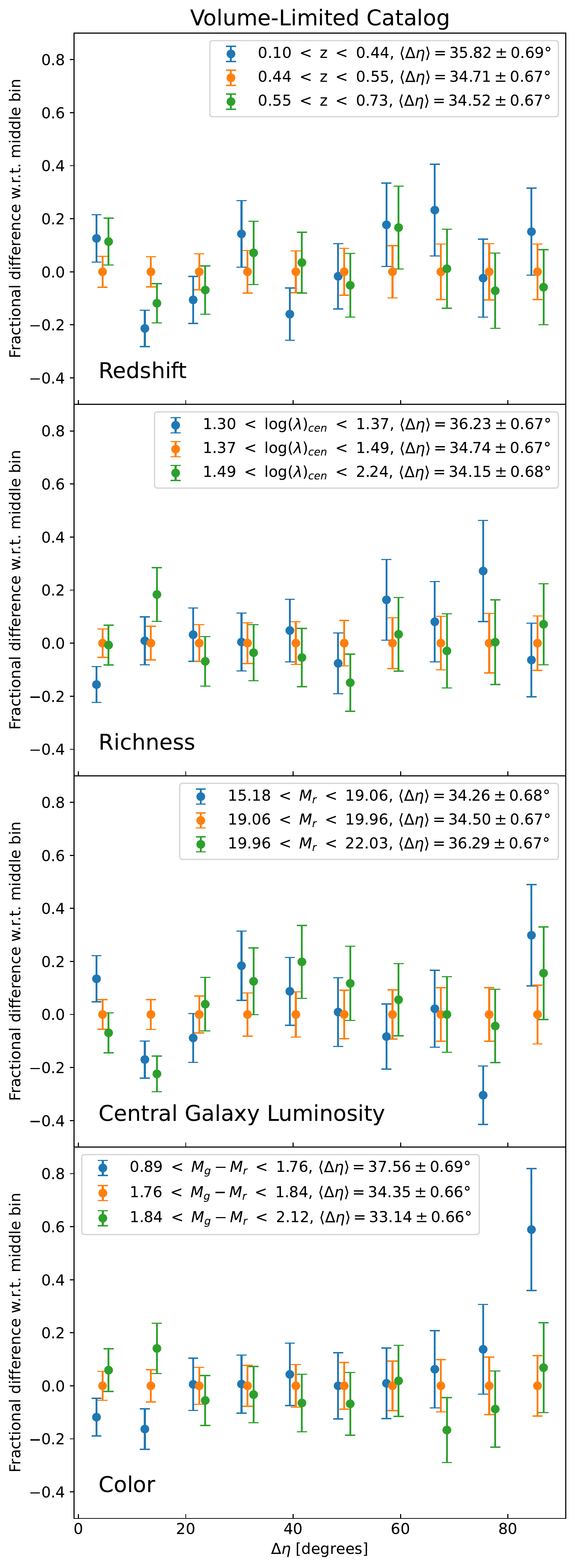}
            \hspace{0.5cm}
            \includegraphics[width=0.45\textwidth]{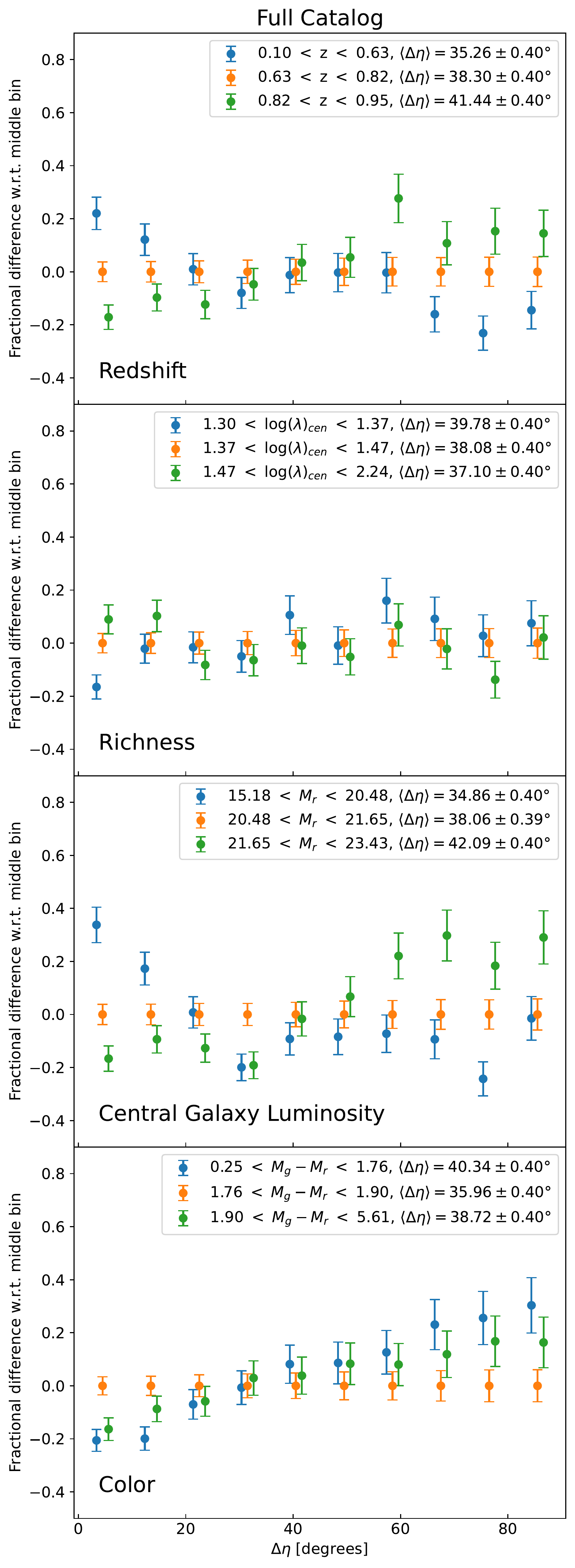}
        \end{center}\vspace{-0.3cm}
        \caption{The position angle difference for clusters split into tertiles of richness, redshift, and central galaxy $r$-band absolute magnitude $M_r$ and $g$-$r$ color. The fractional difference of $\Delta \eta$ with respect to the middle bin is shown. \emph{Left}: Results for the volume-limited catalog. There are very weak indications of trends with the four properties, but only at the 1$\sigma$ level. \emph{Right}: Results for the full catalog, which extends to much higher redshift. There exist highly significant trends in stronger alignment of the central galaxy with the cluster shape when going to higher richness and central galaxy brightness, which are both a proxy for cluster mass. We also find a trend of stronger alignment at lower redshift. These are consistent with the weaker trends in the volume-limited catalog. We also find significant non-monotonic differences in bins of color in the full sample. Points are offset for visibility.
            \label{fig:padiff2}}
    \end{figure*}

    We first compare measurements of the position angle difference $\Delta \eta$, weighted by the probability of satellite galaxies being a cluster member $p_{\mathrm{mem}}$, using the two different methods of measuring $\Delta \eta$ and two estimates of the galaxy shape. Figure \ref{fig:padiff1} shows $\Delta \eta$ for all clusters in the sample, measured by Methods 1 \& 2 and by both \textsc{metacalibration} (MCAL) and im3shape (I3S). In the case of random alignment, we would expect a flat distribution with $\langle \Delta \eta \rangle=45$\textdegree. All four results are generally consistent and show a preference for the alignment of the central galaxy with the overall cluster shape, with the MCAL 2nd moments measurement finding $\langle \Delta \eta \rangle=35.01\pm 0.39$\textdegree, significantly less than 45\textdegree. We find both methods of inferring the cluster satellite distribution shape agree very well cluster-by-cluster, in addition to in the population mean. 

    We are also able to study the dependence of this alignment on both cluster properties (e.g., richness and redshift) and central galaxy properties (e.g., $r$-band absolute magnitude $M_r$ and $g$-$r$ color), which is shown in Fig. \ref{fig:padiff2} for the volume-limited and full cluster catalogs. We split the clusters into tertiles in each of the four quantities, and compare the $\Delta \eta$ distributions. While any possible trends in the volume-limited catalog are very weak (at most the 1$\sigma$ level), we do observe significant trends with the full cluster catalog, which has higher statistical precision and goes to much higher redshift. We find increasing alignment of the central galaxy with the cluster shape for both higher richness clusters and brighter absolute magnitude, as expected, since both are a proxy for cluster mass. We also find a stronger tendency to align for lower redshift clusters, and while there are significant differences in bins of color, there isn't a clear trend in alignment versus color.

    These results are consistent with the weak trends seen in the volume-limited sample. The trends of $\langle \Delta \eta \rangle$ for the full sample are also qualitatively similar to \cite{Huang_2016}, with a slightly better agreement in the low-$z$ tertile selections that better matches the redshift range of the SDSS \textsc{redMaPPer} clusters studied in that paper. In \cite{Huang_2016} they find $\langle \Delta \eta \rangle = 35.07\pm 0.28$\textdegree, while we find $\langle \Delta \eta \rangle = 35.82\pm 0.69$\textdegree, though still extending to higher redshift than the SDSS cluster sample.

    The higher volume probed by the DES data allows us to demonstrate these significant trends across redshift and magnitude for the first time. These results are consistent with a model of the intracluster alignment coalescing as the cluster evolves (at lower redshifts) and being more strongly driven in more massive clusters (larger richness and absolute magnitude). This result would be in conflict with the often-assumed scenario of large-scale alignments of galaxies being frozen in at early times as the galaxies form, and then being disrupted over time. For instance, the typical redshift scaling of analytic IA models (e.g.\ \cite{hirata04,bridle07,2019PhRvD.100j3506B}), assumes this behavior.
    This result, if confirmed with future studies, would provide important insight into how red galaxies align in cluster environments, and potentially with large-scale structure more generally.

    \subsection{Anisotropic distribution of satellite galaxies}

\begin{figure}
        \begin{center}
            \includegraphics[width=\columnwidth]{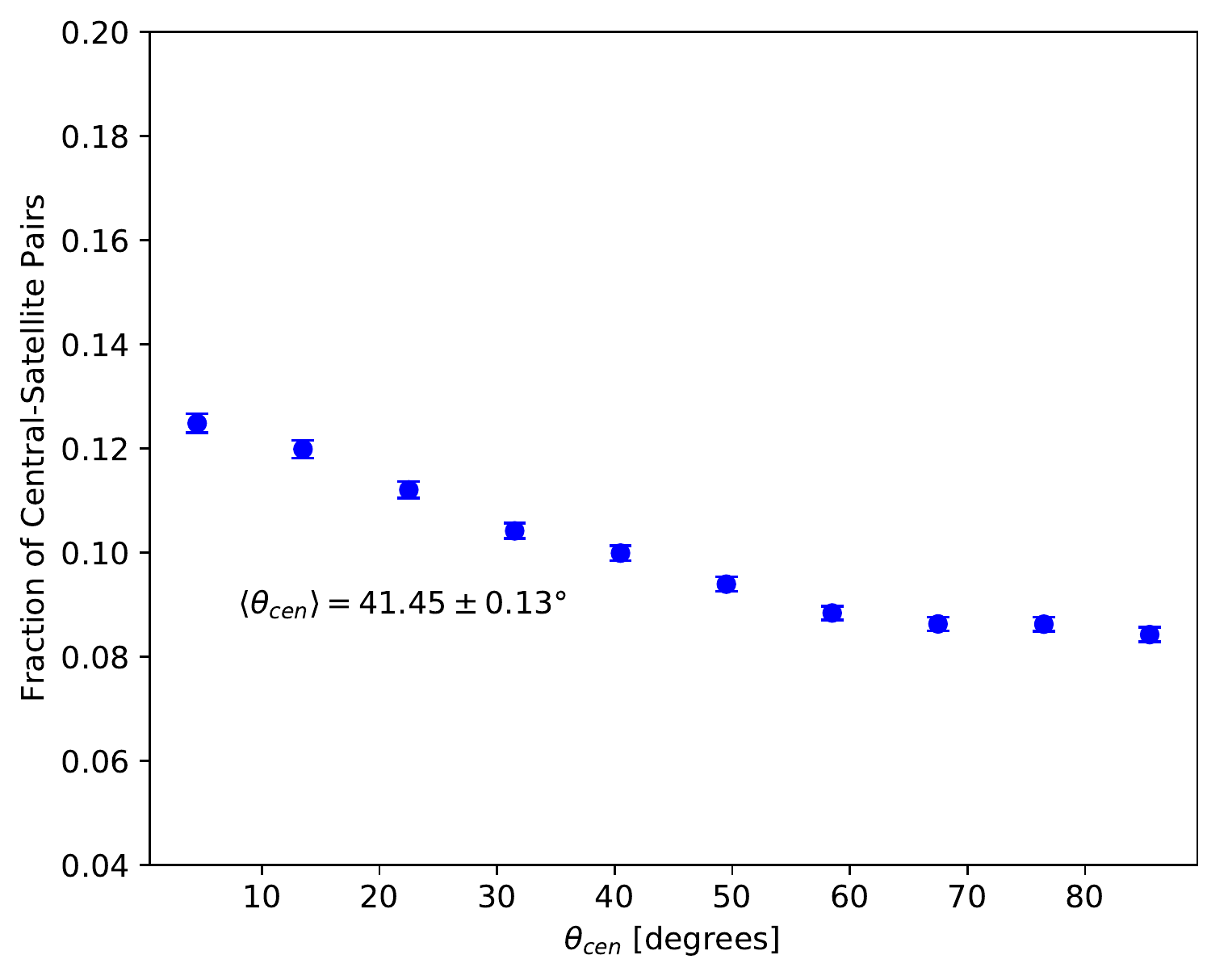}
        \end{center}
        \caption[]{The distribution of the alignment of satellite galaxy positions relative to the position angle of the central galaxy of the cluster ($\theta_{\mathrm{cen}}$). There is a slight preference for satellite galaxies to be aligned closer to the major axis of the central galaxy.
            \label{fig:thetacen}}
    \end{figure}

    Previous studies, including \cite{Huang_2016} of \textsc{redMaPPer} clusters in SDSS, have found a tendency of satellite galaxies to align along the major axis of the central galaxy. We also observe this trend, measured as the distribution of angles $\theta_{\mathrm{cen}}$ weighted by $p_{\mathrm{\mathrm{mem}}}$ between the line connecting central and satellite galaxies with the major axis of the central galaxy. This is shown in Fig. \ref{fig:thetacen}, where we find $\langle  \theta_{\mathrm{cen}}\rangle = 41.45\pm0.13$\textdegree. \updated{The difference in the number of satellites along the major versus minor axes (slope in Fig.  \ref{fig:thetacen}) is perhaps unintuitively much less pronounced than the difference in numbers of clusters with central galaxies aligned vs anti-aligned with the cluster major axis (slope in Fig.  \ref{fig:padiff1}), which is also consistent with what was found in SDSS \textsc{redMaPPer} clusters.
    This is expected, however, since $\theta_{\mathrm{cen}}$ is only the same as $\eta_{\mathrm{cen}}$ in the limit that the cluster ellipticity is 1. Given the model in Sec. \ref{qgsec}, we can model $\theta_{\mathrm{cen}}$ from $\eta_{\mathrm{cen}}$ and find that the two measurements are consistent.}

    \begin{figure}[htp]
        \begin{center}
            \includegraphics[width=0.87\columnwidth]{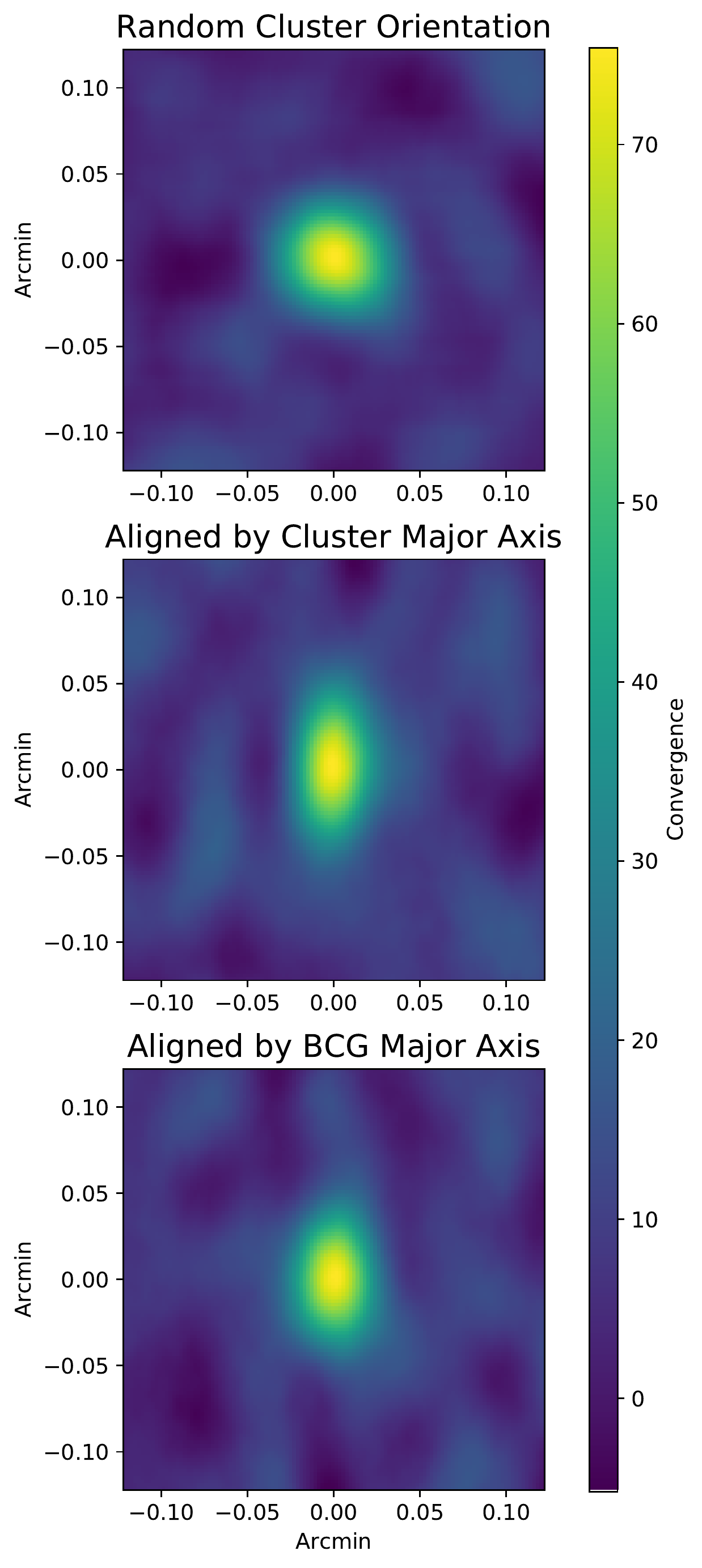}
        \end{center}
        \caption[]{The stacked convergence map centered on the positions of clusters. \emph{Top}: The stacked convergence for clusters with their original orientation on the sky. The measured ellipticity of the mass is consistent with stacked random orientations. \emph{Center}: The stacked convergence for clusters rotated with the position angle inferred from the satellite galaxy distribution oriented vertically. The measured ellipticity of the mass is $e=0.35$.
            \emph{Bottom}: The stacked convergence for clusters rotated with the position angle inferred from the central galaxy major axis. The measured ellipticity of the mass is $e=0.20$.
            \label{fig:mass}}
    \end{figure}

    \subsection{Agreement between halo orientation and galaxy distribution}

    We have used the distribution of satellite galaxies within clusters as a proxy for the shape of the underlying dark matter halo, which is what can be expected to play a major role driving any true intrinsic alignment of the galaxies. To justify this, we compare our cluster shape measurements inferred from the galaxy distribution with the DES Y1 weak lensing convergence `mass' map \citep{2018MNRAS.475.3165C} to confirm the correlation between galaxy satellite distribution and the underlying dark matter halo. The region around each cluster is cut out from the mass map, rotated, and stacked so that the inferred position angle from Sec. \ref{align1} is aligned for all clusters. We show this result in Fig. \ref{fig:mass}, which compares the stacked convergence with original random orientations, which has a nearly isotropic shape, with the cluster stack aligned by position angle, which has a highly anisotropic shape aligned in the direction of the inferred position angle of the stacked clusters.

    The ellipticity inferred from the stacked convergence is $e=0.33$, which agrees well with that inferred from the methods discussed in Sec. \ref{align1}, $e=0.35$. It is important to note that we cannot isolate solely e.g. virially bound galaxies in this process, and it is not clear that all selected cluster members are part of a virially relaxed system (see Sec. \ref{model}). Thus some part of this ellipticity may be incorporating the largest connected filamentary structures near the cluster node in the dark matter distribution.

    We also show in Fig.~\ref{fig:mass} the stacked convergence of clusters oriented by the BCG major axis (see also, for example, \cite{10.1093/mnras/stx3366}; \cite{2020MNRAS.496.2591O,2022MNRAS.513.2178H}). We find this produces a less elliptical stacked signal ($e=0.20$) than orienting by the cluster satellite galaxy distribution in the halo.

    \subsection{Radial alignment of satellite galaxies}\label{radial}

    In addition to the alignment of the central galaxy with the dark matter halo of the cluster, satellite galaxies may also be influenced by the local tidal field, causing a radial alignment of their major axes toward the BCG. We find no evidence for a non-flat distribution, with mean $\phi_{\mathrm{sat}}=44.9\pm 0.8$\textdegree, indicating no statistically significant mean radial alignment of objects between 0\textdegree~and 90\textdegree~within the cluster averaged over all distances from the center. This measurement is weighted to the outer radii of the cluster, where there are more satellite galaxies and could swamp any signal closer to the center of the cluster, where we expect it to exist more strongly due to the cluster halo itself.

\begin{figure}
        \begin{center}
            \includegraphics[width=\columnwidth]{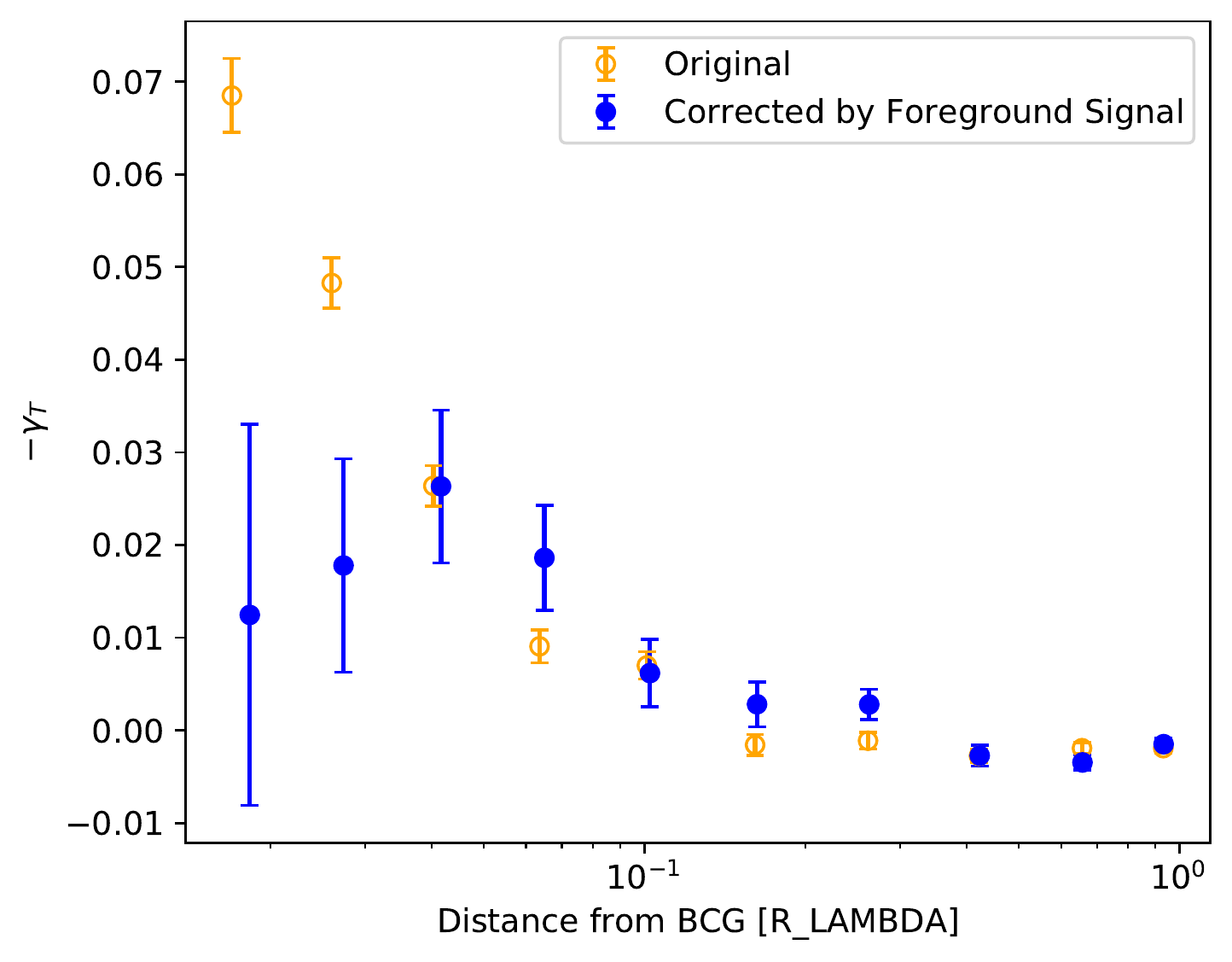}
        \end{center}
        \caption[]{The two-point correlation function $\gamma_{T}$, measuring the mean tangential shape as a function of relative satellite distance from the center of the cluster (negative values indicate radial alignment). The open points are measurements without subtracting the foreground radial alignment signal that we identify as being due to intracluster light impacting the ellipticity measurements of galaxies projected near the center of the cluster. and the solid points are the measurements after subtracting this systematic signal. Within $\sim$0.1$R_\lambda$, there is a significant radial intrinsic alignment signal. The intrinsic alignment signal is consistent with zero on scales larger than $\sim$0.1 $R_\lambda$.
            \label{fig:gammar}}
    \end{figure}

\begin{figure}
        \begin{center}
            \includegraphics[width=\columnwidth]{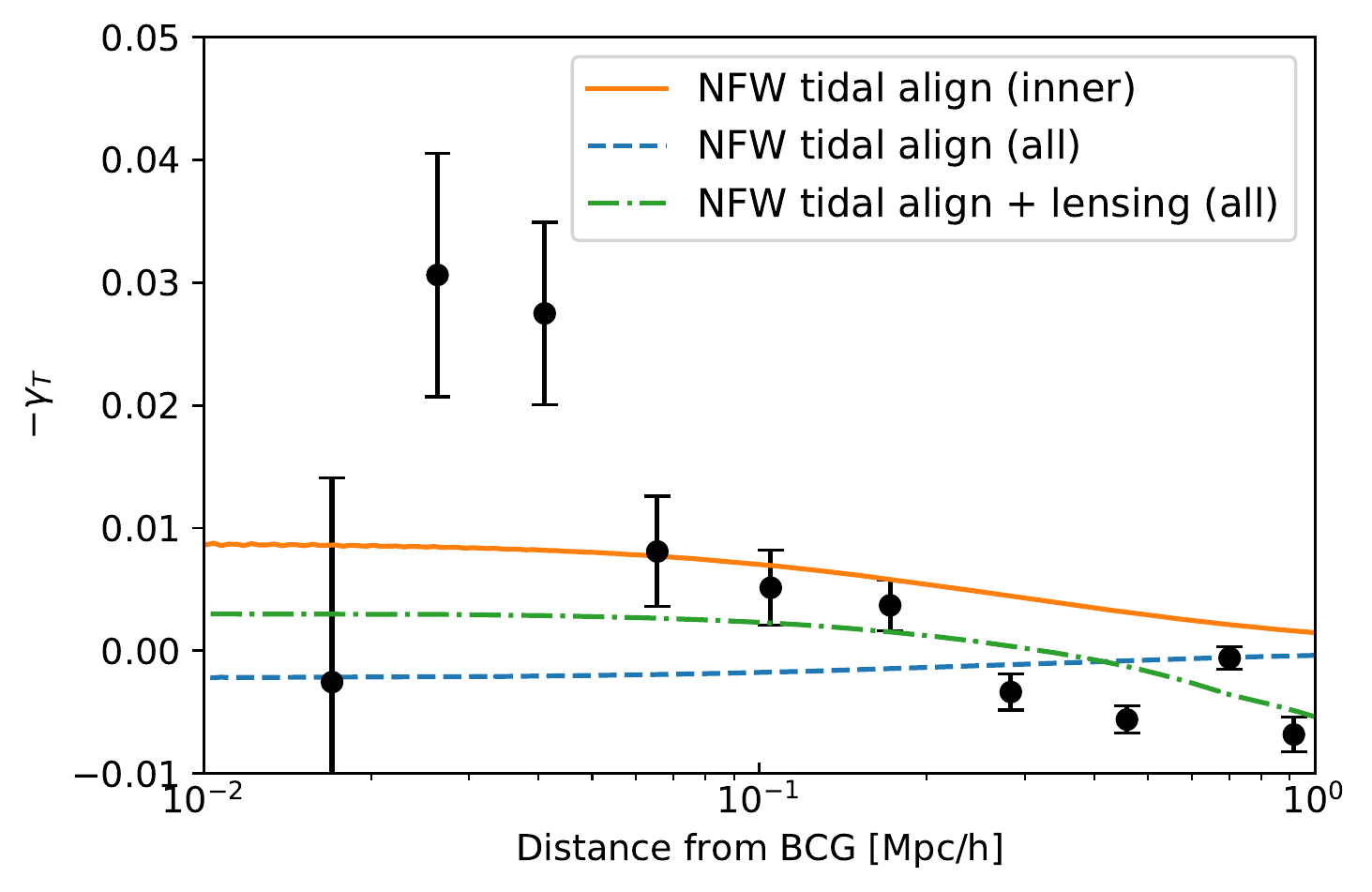}
        \end{center}
        \caption[]{The measured $\gamma_{T}$ signal, corrected for the impact of intracluster light on the ellipticity measurements, in bins of absolute separation. This is compared to the NFW tidal alignment model prediction with $A_{IA}=0.15$ (orange, solid) and $A_{IA}=-0.037$ (blue, dashed), as well as model (green, dash-dot) with both NFW tidal alignment with $A_{IA}=0.06$ and lensing contamination, as described in the text.

            \label{fig:model_fit}}
    \end{figure}

    We also measure the mean shape \cwr{of \textsc{redMaPPer} cluster members} as a function of distance from the cluster center -$\gamma_{T}(R)$, which is shown in Figs. \ref{fig:gammar} \& \ref{fig:model_fit}, with distance from the center of the cluster both as a fraction of the cluster size ($R_{\lambda}$) and in absolute units, respectively. We find a highly significant radial alignment signal within about $0.1 R_{\lambda}$ (or 0.1 Mpc/$h$) of the cluster centers, with a total signal-to-noise $S/N=18$ (``Original'' in Fig. \ref{fig:gammar}). \cwr{In our measurements of $\gamma_{T}(R)$, we apply a ``member boost'' factor to account for the expected (weighted) fraction of cluster members in the sample that are actually foreground/background objects and thus do not contribute to the IA signal. We calculate this factor, \updated{which is a function of distance from the stacked cluster center,} using the \textsc{redMaPPer} membership probabilities $p_{\mathrm{mem}}$: $B_m(r) = \sum_i p_{\mathrm{mem},i} / \sum_i p_{\mathrm{mem},i}^2$. The membership probabilities are also used to weigh each galaxy in the correlation function estimator. This member boost is analogous to the boost factor typically applied to galaxy-galaxy or cluster-galaxy lensing measurements to account for dilution from sources physically associated with the lens.}

\begin{figure}
        \begin{center}
            \includegraphics[width=\columnwidth]{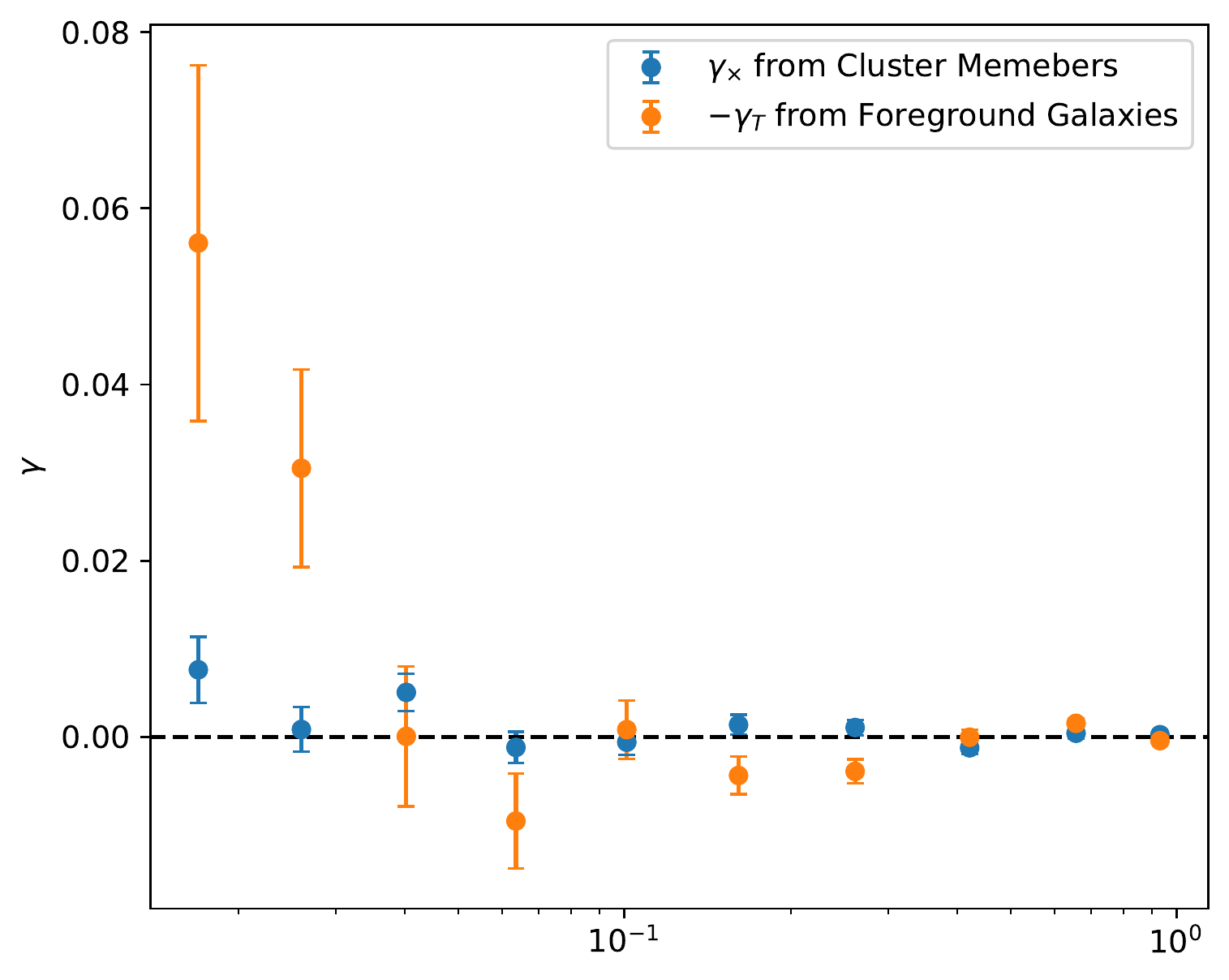}
        \end{center}
        \caption[]{Tests of potential systematic contributions to the measured $\gamma_{T}$ in Fig. \ref{fig:gammar}. The orange dots are the cross-component $\gamma_\times$ using cluster members. The blue dots are the $\gamma_{T}$ signal measured using foreground galaxies around cluster centers. The cross-component should be consistent with zero at the statistical precision of this measurement, and we find that it is. Similarly, since the foreground galaxies are physically disassociated with the local tidal field of the clusters and do not experience lensing due to the clusters, there should also be no physical signal here. We do find evidence of correlation within $\sim$0.05$R_{\lambda}$, which is most likely due to intracluster light near the center of the cluster biasing the shape measurement of overlapping galaxies on the sky. \label{fig:gammax}}
    \end{figure}

    \updated{To test the robustness of this measurement, we also show the result of the $\gamma_\times(R)$ cross-component measurement, which is consistent with zero, in Fig. \ref{fig:gammax}. We also repeat the $\gamma_{T}$ measurement for a sample of galaxies not physically associated with the cluster, but projected in the same line of sight in front of the cluster. This should produce no physical signal, \cwr{as those galaxies are not affected by the potential of the cluster}, yet we find a sharp transition to a significant mean radial alignment within about $0.05 R_{\lambda}$ of the cluster center.
    Previously, \cite{zhangICL} identified an intracluster light profile within DES \textsc{redMaPPer} clusters that is the most plausible cause of this apparent galaxy alignment. The scale of this alignment agrees fairly well with the inner-most profile model component they fit, which may in fact be associated with the edge of the central galaxy profile.}

    \updated{Since we can measure this kind of contamination, we can correct the measured alignment signal for the cluster satellite galaxies by subtracting this foreground signal, which results in the corrected signal shown in Fig. \ref{fig:gammar} in blue. The new covariance for the measurement takes into account the uncertainties from both measurements. All two-point correlation function results will be corrected by this foreground signal. We find that this measured alignment with foreground clusters due to intracluster light is consistent with being unchanged as a function of redshift and richness, so we correct measurements in bins of redshift or richness by the foreground signal for the full cluster population, which has smaller uncertainty.}
    
\begin{figure}
        \begin{center}
            \includegraphics[width=\columnwidth]{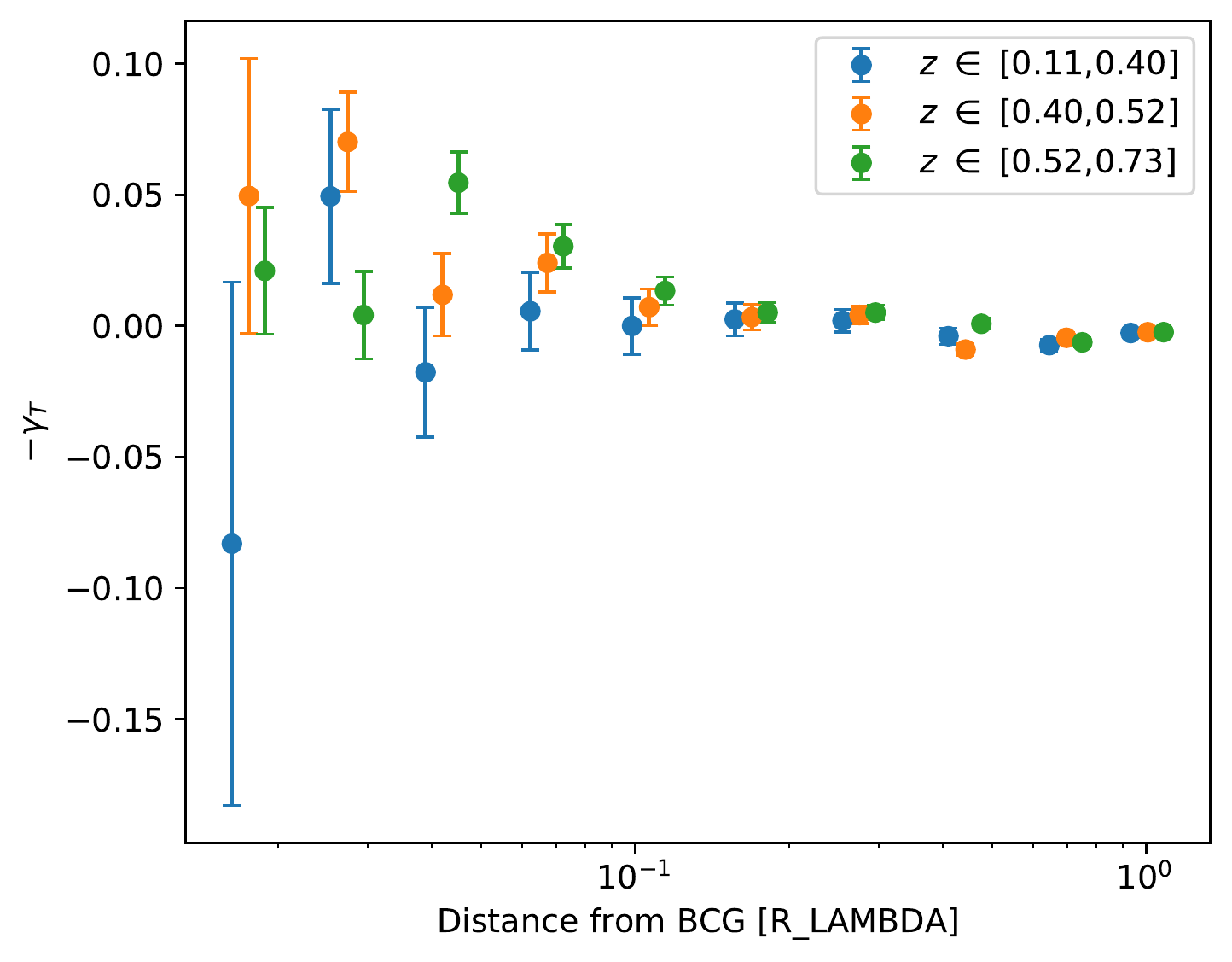}
        \end{center}
        \caption[]{The measured mean radial alignment of satellite galaxies measured for clusters split into three bins of redshift.
            \label{fig:gamma_redshift}}
    \end{figure}

\begin{figure}
        \begin{center}
            \includegraphics[width=\columnwidth]{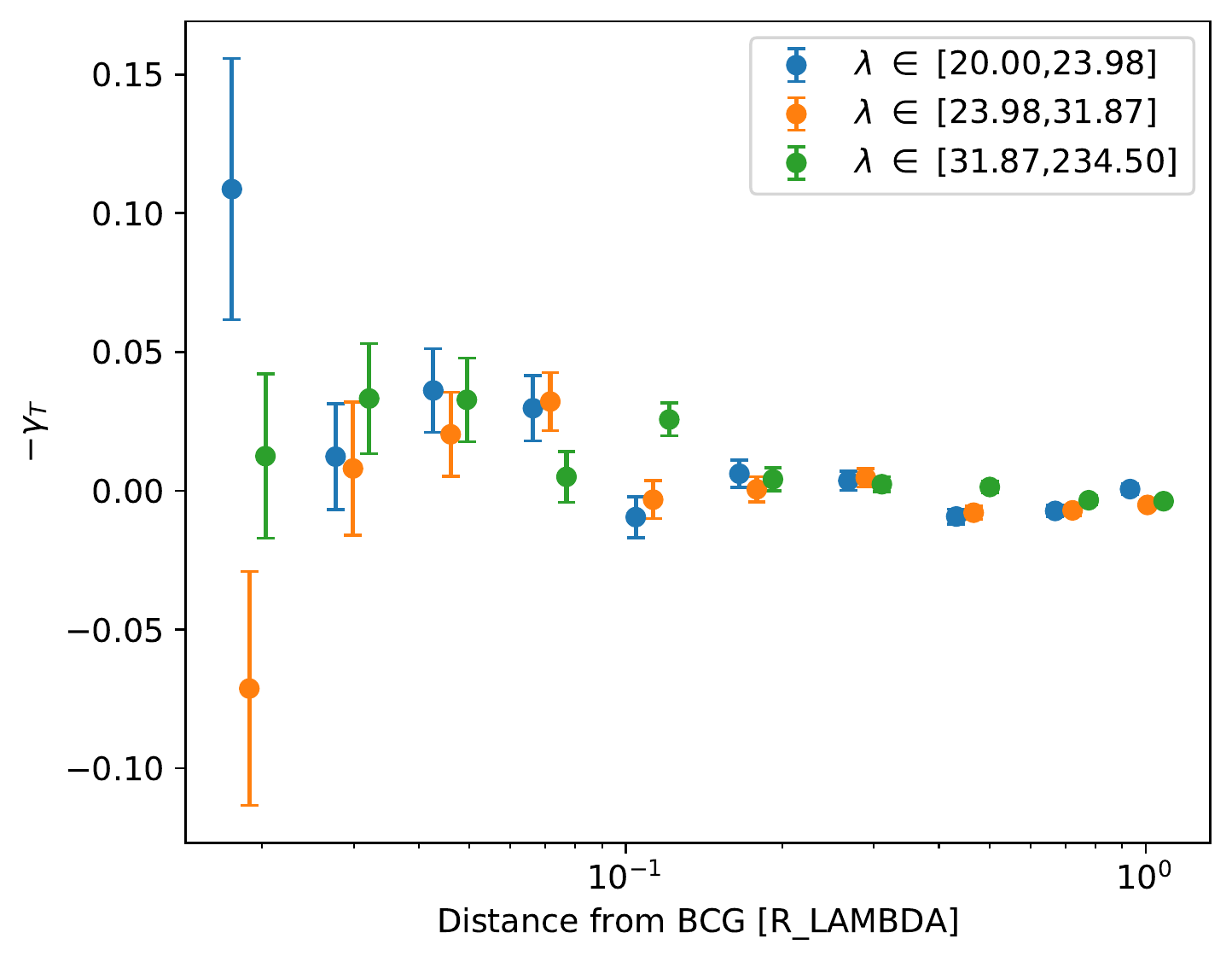}
        \end{center}
        \caption[]{The measured mean radial alignment of satellite galaxies measured for clusters split into three bins of cluster richness.
            \label{fig:gamma_richness}}
    \end{figure}

    \updated{The final radial alignment signal we measure in Figs. \ref{fig:gammar} \& \ref{fig:model_fit} is substantially stronger in amplitude and signal-to-noise than found for the full satellite population with SDSS \textsc{redMaPPer} clusters in \cite{Huang_2016}, with a total signal-to-noise of $\sim$6.
    Given the signal-to-noise of the measurement, we can attempt to look for the evolution of the signal over redshift, shown in Fig. \ref{fig:gamma_redshift}. The radial intrinsic alignment signal from the satellites we observe within $0.1 R_{\lambda}$ of the center of the clusters does have a small indication of portential redshift dependence.  Given recent potential richness-dependent systematics in optical cluster studies \citep{2020PhRvD.102b3509A}, we also consider the richness dependence of the measurement, which is shown in Fig. \ref{fig:gamma_richness}.  We find that the radial intrinsic alignment signal has no richness dependence.}

    \subsection{Impact of measured radial alignment within clusters on cosmology}

    Given the presence of a non-zero radial alignment signal within \textsc{redMaPPer} clusters, it is useful to consider if this signal could leak into estimates of mean tangential shear like $\gamma_t$ or $\Delta\Sigma$. In the cluster lensing measurements in \cite{McClintock:2018bxh}, cosmology is inferred from measurements only at (relative to this study) large scales above 200 kpc, where the alignment signal is small. A buffer in source photometric redshift of 0.1 was also used to remove any sources within $z=0.1$ of the cluster to minimize these effects. However, due to the uncertainty in source redshifts, this leaves a non-zero fraction of cluster members as part of the source catalog. To test any impact of radial alignment leakage, we explicitly remove all cluster members from the source catalog and repeat the measurements in the same bins of richness and redshift from \cite{McClintock:2018bxh}. We find that the impact is much smaller than the uncertainty on the measurement expected even for DES Year 6, indicating this intracluster intrinsic alignment can play no role in systematics of the cluster lensing signal used for cosmological inference. This is due partly to the small fraction of contaminated galaxies and the signal being present most strongly only on scales smaller than those used in the cluster lensing analyses.

    \section{Modeling}\label{model}
    Analytic models of intrinsic alignments typically relate the galaxy shapes to the local tidal field, often in regimes where perturbative approaches are valid (e.g.\ \cite{hirata04,2015JCAP...08..015B,2019PhRvD.100j3506B}. To describe the measured IA signal within \textsc{redMaPPer} clusters, we must in principle include both the fully nonlinear tidal field and nonlinear responses of galaxy shapes to that tidal field. Different approaches have been adopted to treat these effects. A halo model for IA \citep{SB10,2021MNRAS.501.2983F} provides a parameterized description of galaxy shapes and locations within dark matter halos. Similarly, semi-analytic models can be applied to gravity-only simulations to populate dark matter halos with realistically aligned galaxies \citep{2013MNRAS.436..819J,hoffmann22,vanalfeninprep}. These approaches can be compared to both observational data and hydrodynamic simulations (e.g.\ \cite{samuroff21}). However, such comparisons are not yet conclusive, given a combination of small signals and dependence on ``sub-grid'' assumptions.

    In this work, we choose to use a simple nonlinear model to provide an estimate for the expected IA of red galaxies on this scale. Against this estimate, we can then explore the impact of several potential modeling complications relevant on these scales and for galaxy clusters. We believe that these insights can be incorporated into more sophisticated halo modeling in future work.

    \subsection{Nonlinear tidal alignment}
    We start with the ansatz, explored in \cite{2015JCAP...08..015B}, that the IA for red cluster member galaxies can be estimated as proportional to the fully nonlinear tidal field within the cluster. This model is similar in spirit to the ``nonlinear linear alignment'' (NLA) model often used in cosmic shear analyses \cite{hirata04,bridle07,2019MNRAS.489.5453S,kids_ia_cosmology}. However, rather than use the nonlinear dark matter power spectrum, which describes the overall clustering of matter, we use the cluster-matter power spectrum, $P_{\rm cm}$ to calculate the relevant tidal field correlations. As discussed in \cite{2015JCAP...08..015B}, the average galaxy IA, $\gamma_{IA}$ can be described as the (projected) average correlation between the tracer density, in this case galaxy clusters, and the tidal field.

    \begin{align}
        \gamma_{IA} = \frac{1}{2\Pi_{\rm max}}\int_{-\Pi_{\rm max}}^{\Pi_{\rm max}}d\Pi \  \langle \delta_c | \gamma_+\rangle,
    \end{align}
    where $\Pi_{\rm max}$ is the effective projection length. Making the Limber approximation, this expression can be related to $P_{\rm cm}$:
    \begin{align}
        \label{eq:NFW_IA}
        \gamma_{IA} = \frac{1}{2\Pi_{\rm max}}\frac{A_{\rm IA}}{2\pi}
        \int_{0}^{\infty}d\kappa \kappa \ J_2(\kappa r_p) P_{\rm cm}(\kappa),
    \end{align}
    where $A_{\rm IA}$ is the IA amplitude, corresponding to the response of the galaxy shape to the tidal field, and $J_i$ are the (cylindrical) Bessel functions. Finally, for $P_{\rm cm}$, we combine a linear bias model on large scales with an NFW halo contribution \cite{navarro96} on small scales: $P_{cm}= b_c P_{\rm lin} + P_{\rm NFW}$, where $P_{\rm NFW}$ is the Fourier transform of the NFW profile. \updated{We use a bias value of $b_c = 4.27$, a weighted averaged of measurements from \cite{to21}.} On the scales relevant for these intracluster measurements, the NFW contribution dominates over the linear term.

    To generate the NFW profile, we use the mean cluster mass and concentration parameters measured in \cite{McClintock:2018bxh}, corresponding to $M_{200} = 10^{14.1} M_{\odot}$ and $c_{200}=5$.
    We assume a flat $\Lambda$CDM cosmology with $\Omega_m = 0.315$ and $h=0.67$. We note that our results are not sensitive to the assumed cosmological parameters, within reasonable uncertainties.

    As seen in Figure~\ref{fig:model_fit}, the measured data after correcting for the influence of intracluster light are consistent with this fully nonlinear tidal alignment picture, but only on some scales. The positive amplitude measurements (below $\sim200$kpc/$h$ are consistent with the expected tidal alignment, while the negative points on larger scales could be due to contamination from lensing or a different effect not in our model. We discuss several possibilities below.
    When including only scales near the cluster center that exhibit a coherent radial alignment (i.e.~those with the expected IA sign), we find an IA amplitude of $A_{\rm IA} = 0.15 \pm 0.04$ ($\chi^2$/dof = 2.7). This is somewhat smaller than most measurements of the large-scale red galaxy intrinsic alignment amplitude, which tends to be closer to $\sim$1-5, depending on luminosity and details of selection. When fitting the measurements on all scales, we find $A_{\rm IA} = -0.04 \pm 0.02$ ($\chi^2$/dof = 9.4). However, as reflected by the poor fit, this value is mostly a coincidence of tension in mean tangential alignment in the outer regions of the clusters and mean radial alignment in the innermost regions. Alternatively, if we include an additional term, proportional to the \cwr{``member boost'' factor (described above)} which expresses the weighted fraction of non-cluster members, we can allow for lensing contamination in the signal. With this more complex model, we find $A_{\rm IA} = 0.06 \pm 0.03$ ($\chi^2$/dof = 7.1) when fitting all scales. While these models behave qualitatively like our measured alignment signal, only the fit ignoring the outer parts of the cluster have a plausible (though still poor) $\chi^2$ in terms of a probability-to-exceed, with $p=0.02$. This indicates more work is needed to understand the measurements and potential systematics.

    \subsection{Potential limitations to model interpretation}
    We now consider briefly additional effects beyond the measured intracluster light that could potentially impact our interpretation of the comparison of the measured IA and the NFW tidal model. We leave for future work a detailed study of these effects in the context of modeling IA within the one-halo and cluster regime.

    First, the use of the Limber approximation requires an effective line-of-sight projection length that is larger than the transverse separation. While this assumption is typically appropriate for lensing measurements as well as IA measurements that project over $\sim 80-100$~Mpc, it is less clear that the assumption will hold within the 1-halo cluster regime. In particular, because only probable cluster members are selected, the projection length is roughly the same size as the cluster radius. Moreover, if the IA and clustering signals vary considerably within the cluster, the effective projection length will also vary, as it is dominated by the locations of the observed galaxy pairs. As indicated in Eq.~\ref{eq:NFW_IA}, a changing effective projection length will impact the overall normalization of the IA signal. This effect can be understood as follows: as the radial separation decreases, the typical line-of-separation for the counted pairs also decreases, significantly \cwr{increasing the observed average signal}.

    Second, the \textsc{redMaPPer} algorithm selects objects with a membership probability that by construction depends on the distance from the cluster center and provides a weight corresponding to this probability. We use these weights to remove dilution from non cluster members. However, if an appreciable number of galaxies are in fact behind the cluster, this will lead to contamination from gravitational lensing which is not included in our model, which assumes all galaxies are at the cluster redshift.
    Similarly, the membership weights will also alter the effective line-of-sight weighting, e.g.\ compared to Eq.~\ref{eq:NFW_IA}, and we do not take this into account.

    Third, we expect the fraction of cluster members that are fully virialized to increase at smaller radii. If cluster member alignment develops as a response to the local environment during virialization, we would expect the IA signal to increase with the virialized fraction. Conversely, if IA is primarily imprinted by the large-scale tidal field at early times, we may expect the process of virialization to suppress the IA signal. It remains an open question which of these effects dominates IA, both in general and in cluster environments -- see, e.g. \cite{2015JCAP...08..015B,piras18}. However, we note that even assuming a maximal impact of virialization, this would require a very significant change in virialized fraction with radius of the cluster.

    Fourth, our simple ansatz, assuming a fixed linear response to the fully nonlinear field may fail to capture relevant IA physics on these scales. A scale-dependent IA response could capture some of this additional complexity.

    Finally, alignments are measured with respect to an assumed cluster center. Miscentering of \textsc{redMaPPer} clusters (e.g.\ \cite{zhang19,bleem20}) will lead to a suppression of the measured IA signal on the smallest scales. Because $\gtrsim 75\% $ of \textsc{redMaPPer} clusters are well centered \citep{zhang19}, this effect should be subdominant. However, future modeling should account for miscentering for a more precise inference of IA amplitude.

    \section{Conclusions}\label{conclusions}

    As cosmological studies seek to utilize smaller-scale information in the lensing signal, which can contribute significant additional constraining power, it will be key to form a better empirical understanding of the small-scale intrinsic alignment of galaxies. This is particularly true for cluster lensing studies, which probe the most extreme density regions of the universe.  The DES \cwr{Y1} photometric data set is a powerful tool for these studies, due to the large volume probed in which to identify galaxy clusters and the large number of galaxies over that volume with robust shape measurements. The DES \cwr{Y1} redMaPPer cluster catalog extends to nearly $z=1$, providing a wide range of redshift over which to study the evolution of the intrinsic alignment signal in galaxy clusters.

    In this work, we investigate the intracluster alignment of red-sequence galaxies using a variety of metrics that probe: 1) the alignment of the central galaxy with the cluster dark matter halo; 2) the mean distribution and alignment of satellite galaxies with the central galaxy; and 3) the mean radial alignment of satellite galaxies as a function of separation from the cluster center. These are compared across two shape measurement methods, \textsc{metacalibration} and \textsc{im3shape}, and for the full \textsc{redMaPPer} cluster sample and the volume-limited sample used for cosmological inference in DES.

    We find significant trends of alignment in all measurements probed except for the mean alignment of satellite galaxies' position angles relative to the central galaxies in the full populations. We also find that our proxy for the cluster dark matter halo orientation, the distribution of satellite galaxies, agrees well with the orientation of halos inferred by the weak lensing convergence (mass). In particular, we are able to identify significant trends in the alignment of the central galaxy relative to the cluster dark matter halo orientation with increasing cluster richness and central galaxy absolute magnitude (both proxies for cluster mass) and to lower redshifts. This is consistent with an alignment mechanism that increases over time as the cluster evolves, with greater support by more massive clusters, rather than one that is fixed at cluster or galaxy formation and degrades over time with interactions and mergers.

    We are also able to probe the mean radial alignment of cluster satellites relative to the cluster center using the two-point correlation function $\gamma_{T}$, finding a non-zero measurement below 0.2$R_{\lambda}$ or 0.25 Mpc/$h$ with a signal-to-noise of $\sim$6 after correction for systematics in the shape measurements due to intracluster light. Using the full range of scales within the cluster, we find a measurement consistent with zero, due to a tension between the mean radial alignment observed in the inner regions of the clusters and a mean tangential alignment in the outer parts of the clusters. We find both a larger amplitude and higher signal-to-noise than in a previous study of this measurement for \textsc{redMaPPer} clusters in SDSS \citep{Huang_2016,Huang_2017}. The statistical power of this measurement of $\gamma_{T}$ enables us to study its evolution in bins of cluster properties, though we are not able to identify any significant trends with those properties with the current DES Year 1 data set.

    The statistical power of these kinds of radial alignment measurements in cluster regions can enable new constraints on simulations and models of small-scale intrinsic alignment behavior. We make a first attempt to compare the measurement to a simple tidal intrinsic alignment model inferred from the constraints on the NFW halo profile for these \textsc{redMaPPer} clusters, and find an alignment amplitude $A_{\textrm{IA}}=0.15 \pm 0.04$ ($p=0.02$)  when excluding data near the edge of the cluster. We discuss several potential caveats with this simple modeling approach and leave a more extensive attempt to model or simulate the measurement to future works.

    \updated{The intrinsic alignment of galaxies in the one-halo regime has implications for cosmic shear measurements. Previous studies have considered this impact, e.g.\ \cite{2015A&A...575A..48S,2021MNRAS.501.2983F}, finding that the impact is likely significant, but with a large uncertainty due to the unknown degree of alignments and their dependence on halo mass. In probing alignments at the cluster mass scale with good precision, our measurement will allow these predictions to be made with greater certainty. We leave these calculations for future work but note that our measurements indicate IA that may be somewhat larger than what is assumed in the forecast of \citep{2015A&A...575A..48S}.}

    The measurements of intracluster intrinsic alignment of red-sequence galaxies presented here are just an example of the power available in large photometric data sets like DES to study intrinsic alignment phenomena. We have used here the first year of DES data, which only covers one-third of the full survey area to half image depth. We expect significant increases in statistical power for these studies in the full DES data set and future surveys like Euclid, the Vera C. Rubin Observatory Legacy Survey of Space and Time, and the Roman Space Telescope. These future measurements will unlock new potential for constraining small-scale astrophysics to inform more robust cosmological analyses with lensing.

    \section*{Acknowledgements}

    MT is supported by DOE Award SC0000253548.
    JB is supported by NSF Award AST-2206563.
    We acknowledge use of the lux supercomputer at UC Santa Cruz, funded by NSF MRI grant AST 1828315.

    Funding for the DES Projects has been provided by the U.S. Department of Energy, the U.S. National Science Foundation, the Ministry of Science and Education of Spain,
    the Science and Technology Facilities Council of the United Kingdom, the Higher Education Funding Council for England, the National Center for Supercomputing
    Applications at the University of Illinois at Urbana-Champaign, the Kavli Institute of Cosmological Physics at the University of Chicago,
    the Center for Cosmology and Astro-Particle Physics at the Ohio State University,
    the Mitchell Institute for Fundamental Physics and Astronomy at Texas A\&M University, Financiadora de Estudos e Projetos,
    Funda{\c c}{\~a}o Carlos Chagas Filho de Amparo {\`a} Pesquisa do Estado do Rio de Janeiro, Conselho Nacional de Desenvolvimento Cient{\'i}fico e Tecnol{\'o}gico and
    the Minist{\'e}rio da Ci{\^e}ncia, Tecnologia e Inova{\c c}{\~a}o, the Deutsche Forschungsgemeinschaft and the Collaborating Institutions in the Dark Energy Survey.

    The Collaborating Institutions are Argonne National Laboratory, the University of California at Santa Cruz, the University of Cambridge, Centro de Investigaciones Energ{\'e}ticas,
    Medioambientales y Tecnol{\'o}gicas-Madrid, the University of Chicago, University College London, the DES-Brazil Consortium, the University of Edinburgh,
    the Eidgen{\"o}ssische Technische Hochschule (ETH) Z{\"u}rich,
    Fermi National Accelerator Laboratory, the University of Illinois at Urbana-Champaign, the Institut de Ci{\`e}ncies de l'Espai (IEEC/CSIC),
    the Institut de F{\'i}sica d'Altes Energies, Lawrence Berkeley National Laboratory, the Ludwig-Maximilians Universit{\"a}t M{\"u}nchen and the associated Excellence Cluster Universe,
    the University of Michigan, NFS's NOIRLab, the University of Nottingham, The Ohio State University, the University of Pennsylvania, the University of Portsmouth,
    SLAC National Accelerator Laboratory, Stanford University, the University of Sussex, Texas A\&M University, and the OzDES Membership Consortium.

    Based in part on observations at Cerro Tololo Inter-American Observatory at NSF’s NOIRLab (NOIRLab Prop. ID 2012B-0001; PI: J. Frieman), which is managed by the Association of Universities for Research in Astronomy (AURA) under a cooperative agreement with the National Science Foundation.

    The DES data management system is supported by the National Science Foundation under Grant Numbers AST-1138766 and AST-1536171.
    The DES participants from Spanish institutions are partially supported by MICINN under grants ESP2017-89838, PGC2018-094773, PGC2018-102021, SEV-2016-0588, SEV-2016-0597, and MDM-2015-0509, some of which include ERDF funds from the European Union. IFAE is partially funded by the CERCA program of the Generalitat de Catalunya.
    Research leading to these results has received funding from the European Research
    Council under the European Union's Seventh Framework Program (FP7/2007-2013) including ERC grant agreements 240672, 291329, and 306478.
    We  acknowledge support from the Brazilian Instituto Nacional de Ci\^encia
    e Tecnologia (INCT) e-Universe (CNPq grant 465376/2014-2).

    This manuscript has been authored by Fermi Research Alliance, LLC under Contract No. DE-AC02-07CH11359 with the U.S. Department of Energy, Office of Science, Office of High Energy Physics.

\section*{Data Availability}
The data underlying this article will be shared on reasonable request to the corresponding author.




    \bibliographystyle{mnras}
    \bibliography{short} 

    \section*{Affiliations}
$^{1}$ Department of Physics, Duke University Durham, NC 27708, USA\\
$^{2}$ Physics Department, UC Santa Cruz, 1156 High Street, Santa Cruz, CA 95064\\
$^{3}$ Department of Physics, Northeastern University, Boston, MA 02115, USA\\
$^{4}$ Institute of Cosmology and Gravitation, University of Portsmouth, Portsmouth, PO1 3FX, UK\\
$^{5}$ Argonne National Laboratory, 9700 South Cass Avenue, Lemont, IL 60439, USA\\
$^{6}$ Department of Astronomy and Astrophysics, University of Chicago, Chicago, IL 60637, USA\\
$^{7}$ Kavli Institute for Cosmological Physics, University of Chicago, Chicago, IL 60637, USA\\
$^{8}$ Astronomy Unit, Department of Physics, University of Trieste, via Tiepolo 11, I-34131 Trieste, Italy\\
$^{9}$ INAF-Osservatorio Astronomico di Trieste, via G. B. Tiepolo 11, I-34143 Trieste, Italy\\
$^{10}$ Institute for Fundamental Physics of the Universe, Via Beirut 2, 34014 Trieste, Italy\\
$^{11}$ Lawrence Berkeley National Laboratory, 1 Cyclotron Road, Berkeley, CA 94720, USA\\
$^{12}$ University Observatory, Faculty of Physics, Ludwig-Maximilians-Universit\"at, Scheinerstr. 1, 81679 Munich, Germany\\
$^{13}$ Fermi National Accelerator Laboratory, P. O. Box 500, Batavia, IL 60510, USA\\
$^{14}$ Center for Astrophysical Surveys, National Center for Supercomputing Applications, 1205 West Clark St., Urbana, IL 61801, USA\\
$^{15}$ Department of Astronomy, University of Illinois at Urbana-Champaign, 1002 W. Green Street, Urbana, IL 61801, USA\\
$^{16}$ Department of Physics and Astronomy, University of Pennsylvania, Philadelphia, PA 19104, USA\\
$^{17}$ Department of Applied Mathematics and Theoretical Physics, University of Cambridge, Cambridge CB3 0WA, UK\\
$^{18}$ Department of Physics, University of Arizona, Tucson, AZ 85721, USA\\
$^{19}$ Department of Astrophysical Sciences, Princeton University, Peyton Hall, Princeton, NJ 08544, USA\\
$^{20}$ Institut d'Estudis Espacials de Catalunya (IEEC), 08034 Barcelona, Spain\\
$^{21}$ Institute of Space Sciences (ICE, CSIC),  Campus UAB, Carrer de Can Magrans, s/n,  08193 Barcelona, Spain\\
$^{22}$ Kavli Institute for Particle Astrophysics \& Cosmology, P. O. Box 2450, Stanford University, Stanford, CA 94305, USA\\
$^{23}$ SLAC National Accelerator Laboratory, Menlo Park, CA 94025, USA\\
$^{24}$ Brookhaven National Laboratory, Bldg 510, Upton, NY 11973, USA\\
$^{25}$ Department of Physics and Astronomy, Stony Brook University, Stony Brook, NY 11794, USA\\
$^{26}$ Instituto de Astrofisica de Canarias, E-38205 La Laguna, Tenerife, Spain\\
$^{27}$ Laborat\'orio Interinstitucional de e-Astronomia - LIneA, Rua Gal. Jos\'e Cristino 77, Rio de Janeiro, RJ - 20921-400, Brazil\\
$^{28}$ Universidad de La Laguna, Dpto. AstrofÃ­sica, E-38206 La Laguna, Tenerife, Spain\\
$^{29}$ Centro de Investigaciones Energ\'eticas, Medioambientales y Tecnol\'ogicas (CIEMAT), Madrid, Spain\\
$^{30}$ Institute for Astronomy, University of Edinburgh, Edinburgh EH9 3HJ, UK\\
$^{31}$ Excellence Cluster Origins, Boltzmannstr.\ 2, 85748 Garching, Germany\\
$^{32}$ Max Planck Institute for Extraterrestrial Physics, Giessenbachstrasse, 85748 Garching, Germany\\
$^{33}$ Universit\"ats-Sternwarte, Fakult\"at f\"ur Physik, Ludwig-Maximilians Universit\"at M\"unchen, Scheinerstr. 1, 81679 M\"unchen, Germany\\
$^{34}$ Department of Astronomy, University of Michigan, Ann Arbor, MI 48109, USA\\
$^{35}$ Department of Physics, University of Michigan, Ann Arbor, MI 48109, USA\\
$^{36}$ Institute of Astronomy, University of Cambridge, Madingley Road, Cambridge CB3 0HA, UK\\
$^{37}$ Kavli Institute for Cosmology, University of Cambridge, Madingley Road, Cambridge CB3 0HA, UK\\
$^{38}$ CNRS, UMR 7095, Institut d'Astrophysique de Paris, F-75014, Paris, France\\
$^{39}$ Sorbonne Universit\'es, UPMC Univ Paris 06, UMR 7095, Institut d'Astrophysique de Paris, F-75014, Paris, France\\
$^{40}$ Department of Physics \& Astronomy, University College London, Gower Street, London, WC1E 6BT, UK\\
$^{41}$ School of Mathematics and Physics, University of Queensland,  Brisbane, QLD 4072, Australia\\
$^{42}$ Department of Physics, IIT Hyderabad, Kandi, Telangana 502285, India\\
$^{43}$ Jet Propulsion Laboratory, California Institute of Technology, 4800 Oak Grove Dr., Pasadena, CA 91109, USA\\
$^{44}$ Institute of Theoretical Astrophysics, University of Oslo. P.O. Box 1029 Blindern, NO-0315 Oslo, Norway\\
$^{45}$ Santa Cruz Institute for Particle Physics, Santa Cruz, CA 95064, USA\\
$^{46}$ Center for Cosmology and Astro-Particle Physics, The Ohio State University, Columbus, OH 43210, USA\\
$^{47}$ Department of Physics, The Ohio State University, Columbus, OH 43210, USA\\
$^{48}$ Center for Astrophysics $\vert$ Harvard \& Smithsonian, 60 Garden Street, Cambridge, MA 02138, USA\\
$^{49}$ Australian Astronomical Optics, Macquarie University, North Ryde, NSW 2113, Australia\\
$^{50}$ Lowell Observatory, 1400 Mars Hill Rd, Flagstaff, AZ 86001, USA\\
$^{51}$ Departamento de F\'isica Matem\'atica, Instituto de F\'isica, Universidade de S\~ao Paulo, CP 66318, S\~ao Paulo, SP, 05314-970, Brazil\\
$^{52}$ George P. and Cynthia Woods Mitchell Institute for Fundamental Physics and Astronomy, and Department of Physics and Astronomy, Texas A\&M University, College Station, TX 77843,  USA\\
$^{53}$ Instituci\'o Catalana de Recerca i Estudis Avan\c{c}ats, E-08010 Barcelona, Spain\\
$^{54}$ Institut de F\'{\i}sica d'Altes Energies (IFAE), The Barcelona Institute of Science and Technology, Campus UAB, 08193 Bellaterra (Barcelona) Spain\\
$^{55}$ Department of Physics, Carnegie Mellon University, Pittsburgh, Pennsylvania 15312, USA\\
$^{56}$ Observat\'orio Nacional, Rua Gal. Jos\'e Cristino 77, Rio de Janeiro, RJ - 20921-400, Brazil\\
$^{57}$ Department of Physics, University of Genova and INFN, Via Dodecaneso 33, 16146, Genova, Italy\\
$^{58}$ Department of Physics and Astronomy, Pevensey Building, University of Sussex, Brighton, BN1 9QH, UK\\
$^{59}$ School of Physics and Astronomy, University of Southampton,  Southampton, SO17 1BJ, UK\\
$^{60}$ Computer Science and Mathematics Division, Oak Ridge National Laboratory, Oak Ridge, TN 37831\\
$^{61}$ Waldorf High School, Belmont, MA 02478






\bsp	
\label{lastpage}
\end{document}